\documentclass[aps,prb,showpacs, twocolumn,superscriptaddress]{revtex4-1}

\usepackage{bm}
\usepackage{amssymb}
\usepackage{amsmath}
\usepackage{latexsym}
\usepackage{graphicx}
\usepackage{placeins}
\usepackage{booktabs}
\usepackage{subfig}
\usepackage{multirow}
\usepackage{color}
\usepackage{algorithm}
\usepackage{algorithmic}

\DeclareMathOperator{\Tr}{Tr}

\DeclareMathOperator{\diag}{diag}

\newcommand{\ie}{\textit{i.e.}{}}

\newcommand{\ud}{\,\mathrm{d}}

\newcommand{\Or}{\mathcal{O}}

\newcommand{\mc}[1]{\mathcal{#1}}

\newcommand{\abs}[1]{\left\lvert#1\right\rvert}

\newcommand{\averageop}[3]{\langle#1\lvert#2\rvert#3\rangle}

\newcommand{\braket}[2]{\langle#1\vert#2\rangle}

\newcommand{\xc}{{\mathrm{xc}}}

\newcommand{\tot}{\mathrm{tot}}

\renewcommand{\Im}{\mathfrak{Im}}

\newcommand{\barint}{\kern4pt \raise3.4pt\hbox{\vrule height.6pt
    width7pt} \kern-11pt \int}

\newcommand{\Hnnz}{ H_{\mathrm{nnz}}\% }
\newcommand{\Lnnz}{ L_{\mathrm{nnz}}\% }

\begin{document}


\title{Accelerating Atomic Orbital-based Electronic Structure Calculation via Pole Expansion and Selected Inversion}

\author{Lin Lin}
\affiliation{Computational Research Division, Lawrence Berkeley National Laboratory, Berkeley, CA 94720}
\author{Mohan Chen}
\affiliation{Key Laboratory of Quantum Information, CAS, University of Science and Technology of China, Hefei, Anhui 230026, P. R. China}
\author{Chao Yang}
\affiliation{Computational Research Division, Lawrence Berkeley National Laboratory, Berkeley, CA 94720}
\author{Lixin He}
\affiliation{Key Laboratory of Quantum Information, CAS, University of Science and Technology of China, Hefei, Anhui 230026, P. R. China}

\begin{abstract}
	We describe how to apply the recently developed pole expansion and selected inversion (PEXSI) technique to Kohn-Sham density function theory (DFT) electronic structure calculations that are based on atomic orbital discretization.  We give analytic expressions for evaluating the charge density, the total energy, the Helmholtz free energy and the atomic forces (including both the Hellman-Feynman force and the Pulay force) without using the eigenvalues and eigenvectors of the Kohn-Sham Hamiltonian.  We also show how to update the chemical potential without using Kohn-Sham eigenvalues.  The advantage of using PEXSI is that it has a much lower computational complexity than that associated with the matrix diagonalization procedure.  We demonstrate the performance gain by comparing the timing of PEXSI with that of diagonalization on insulating and metallic nanotubes.  For these quasi-1D systems, the complexity of PEXSI is linear with respect to the number of atoms.  This linear scaling can be observed in our computational experiments when the number of atoms in a nanotube is larger than a few hundreds.  Both the wall clock time and the memory requirement of PEXSI is modest.  This makes it even possible to perform Kohn-Sham DFT calculations for 10,000-atom nanotubes with a sequential implementation of the selected inversion algorithm.  We also perform an accurate geometry optimization calculation on a truncated (8,0) boron-nitride nanotube system containing 1024 atoms.  Numerical results indicate that the use of PEXSI does not lead to loss of accuracy required in a practical DFT calculation.
\end{abstract}


\maketitle

\section{Introduction}
Electronic structure calculations based on solving the Kohn-Sham
density functional theory (KSDFT) play an important role in the analysis of electronic,
structural and optical properties of molecules, solids and other
nano structures. The efficiency of such a calculation depends
largely on the computational cost associated with the evaluation of the
electron charge density for a given potential within a self-consistent
field (SCF) iteration.  The most straightforward way to perform such an
evaluation is to partially diagonalize the Kohn-Sham Hamiltonian by
computing a set of eigenvectors corresponding to the algebraically
smallest eigenvalues of the Hamiltonian. The complexity of
this approach is $\mathcal{O}(N_e^3)$, where $N_e$ is the number of
electrons in the atomistic system of interest.  As the number of atoms
or electrons in the system increases, the cost of diagonalization
becomes prohibitively expensive.

Linear scaling algorithms (or $\mathcal{O}(N_e)$ scaling
methods, see for
example~\cite{BowlerMiyazakiGillan2002,FattebertBernholc2000,HineHaynesMostofiEtAl2009,Yang1991,LiNunesVanderbilt1993,McWeeny1960},
and review articles~\cite{Goedecker1999,BowlerMiyazaki2012})
are attractive alternatives for solving KSDFT.  The traditional linear scaling
methods use the nearsightedness principle, which asserts that the density
perturbation induced by a local change in the external potential decays
exponentially away from where the perturbation is applied. Consequently,
the off-diagonal elements of the density matrix decay
exponentially away from the diagonal~\cite{Kohn1996,ProdanKohn2005}.
Strictly speaking, the nearsightedness property is valid for insulating
systems but not for metallic systems.

In order to design a fast algorithm that is accurate for
both insulating and metallic systems, we use an
equivalent formulation of KSDFT, in which the charge density is
evaluated as the diagonal of the Fermi-Dirac function evaluated at a
fixed Kohn-Sham Hamiltonian.  By approximating the Fermi-Dirac function
through a pole expansion technique~\cite{LinLuYingE2009}, we can reduce
the problem of computing the charge density to that of computing the
diagonal of the inverses of a number of shifted Kohn-Sham Hamiltonians.
This approach was pursued by a number of researchers in the past. The
cost of this approach depends on the number of poles required to expand
the Fermi-Dirac function and the cost for computing the diagonal of the
inverse of a shifted Kohn-Sham Hamiltonian.

The recent work by Lin et al.~\cite{LinLuYingE2009} provides an accurate and
efficient pole-expansion scheme for approximating the Fermi-Dirac function.
The number of poles required in this approach is proportional to
$\log (\beta \Delta E)$, where $\beta$ is proportional to the inverse
of the temperature, and $\Delta E$ is the spectral width
of the Kohn-Sham Hamiltonian. (i.e. the difference between the largest
and the smallest eigenvalues). This number of expansion terms, or the
pole count here is significantly lower than those given in the previous
approaches~\cite{BaroniGiannozzi1992,Goedecker1993,Ozaki2007,parrinello2008,Ozaki2010}.
When temperature decreases, $\beta$ becomes large. The favorable
scaling of the pole expansion allows us to treat both insulating and
metallic systems efficiently at room temperature or even lower
temperature.

Furthermore, an efficient selected inversion algorithm for computing the
inverse of the diagonal of a shifted Kohn-Sham Hamiltonian without
computing the full inverse of the Hamiltonian has been developed
~\cite{LinLuYingCarE:09,LinYangMezaEtAl2011,LinYangLuEtAl2011}.  
The idea of using the inverse of shifted Hamiltonian operator
(Green's function) for reducing the complexity of Kohn-Sham density
functional theory has also been pursued in other
recent works~\cite{Varga2010,Ozaki2010}. In the selected inversion method,
the complexity of this algorithm is $\mathcal{O}(N_e)$ for quasi-1D systems
such as nanorods, nanotubes and nanowires, $\mathcal{O}(N_e^{3/2})$ for
quasi-2D systems such as graphene and surfaces, and $\mathcal{O}(N_e^2)$
for 3D bulk systems.   In exact arithmetic, the selected inversion
algorithm gives the exact diagonal of the inverse, i.e., the algorithm
does not rely on any type of localization or truncation scheme.  For
insulating systems, the use of localization and truncation can be
combined with selected inversion to reduce the complexity of the
algorithm further to $\mathcal{O}(N_e)$ even for general 3D systems.

In the previous work~\cite{LinYangMezaEtAl2011,LinYangLuEtAl2011},
we used the pole expansion and selected inversion (PEXSI) technique
to solve the Kohn-Sham problem discretized by
a finite difference scheme.  However, it is worth pointing out
that PEXSI is a general technique that is not limited to discretized
problems obtained from finite difference.  In particular,
it can be readily applied to discretized Kohn-Sham problems
obtained from any localized basis expansion technique.  In this paper,
we describe how PEXSI can be used to speed up the solution of a discretized
Kohn-Sham problem obtained from an atomic orbital basis expansion.  We
show that electron charge density, total energy, Helmholtz free energy
and atomic forces can all be efficiently calculated by using PEXSI.

We demonstrate the performance gain we can achieve by comparing
PEXSI with the LAPACK diagonalization subroutine {\tt dsygv} on two
types of nanotubes. We show that by using the PEXSI technique, it is
possible to perform electronic structure calculations accurately for a nanotube
that contains 10,000 atoms with a sequential implementation of
the selected inversion algorithm within a reasonable
amount of time. This is not possible with the sequential LAPACK
subroutine.  For this example, PEXSI exhibits linear scaling when the
system size exceeds a few hundred atoms.


This paper is organized as follows.
In section~\ref{sec:theory}, we show how the PEXSI technique previously
developed~\cite{LinLuYingE2009,LinLuYingCarE:09,LinYangMezaEtAl2011,LinYangLuEtAl2011} can be
extended to solve discretized Kohn-Sham problems obtained from an
atomic orbital expansion scheme. In particular, we will
show how charge density, total energy, free energy and force can be calculated
in this formalism.  We will also discuss how to update the chemical
potential.  In section~\ref{sec:examples}, we report the performance of
PEXSI on two quasi-1D test problems.

Throughout the paper, we use $\Im(A)$ to denote the imaginary part of a
complex matrix $A$. A properly defined inner product between two
functions $f$ and $g$ is sometimes denoted by $\langle f \vert g\rangle$.
The diagonal of a matrix $A$
is sometimes denoted by $\mbox{diag}(A)$. We use $\hat{H}(x,x')$ to
denote the Hamiltonian operator, and $H, S$ to denote the discretized
Hamiltonian matrix and the corresponding overlap matrix obtained
from a basis set $\Phi$.
Similarly $\hat{\gamma}(x,x')$ denotes the single particle density
matrix operator, and the corresponding electron density
is denoted by $\hat{\rho}(x)$.  The matrix $\Gamma$ denotes the 
single particle density matrix represented under a basis set $\Phi$.
It will be used to define the electron density $\hat{\rho}$ and
the total energy $E_{\tot}$.  In a finite temperature \textit{ab initio}
molecular dynamics simulation, we also need the Helmholtz free
energy
$\mc{F}_{\tot}$, and the atomic forces on the nuclei $\{F_{I}\}$.  To
compute these quantities without using Kohn-Sham eigenvalues and
Kohn-Sham orbitals, we need the free energy density matrix
$\Gamma^{\mc{F}}$
and the energy density matrix $\Gamma^{E}$. In PEXSI, these
matrices are approximated by a finite $P$-term pole
expansion, denoted by $\Gamma_{P},\Gamma^{\mc{F}}_{P},\Gamma^{E}_{P}$
respectively.  However, to simplify notation, we will drop the
subscript $P$ and simply use $\Gamma,\Gamma^{\mc{F}},\Gamma^{E}$ to
denote the approximated matrices unless otherwise noted.

%
%
%
%
%

\section{Theory} \label{sec:theory}
The ground-state electron charge density $\hat{\rho}(x)$ of an atomistic
system can be obtained from the self-consistent solution to
the Kohn-Sham equations
\begin{equation}
  \hat{H}\left[\hat{\rho}(x)\right] \psi_i(x) = \psi_i(x) \varepsilon_i,
\label{kseqs}
\end{equation}
where $\hat{H}$ is the Kohn-Sham Hamiltonian that depends on $\hat{\rho}(x)$,
$\{\psi_i(x)\}$ are the Kohn-Sham orbitals that
satisfy the orthonormality constraints
\begin{equation}
 \int \psi^{\ast}_i(x) \psi_j(x) dx = \delta_{ij},
  \label{orth}
\end{equation}
and the eigenvalue $\varepsilon_i$ is often known as the $i$th Kohn-Sham
energy level.  Using the Kohn-Sham orbitals,
we can define the charge density by
\begin{equation}
  \hat{\rho}(x) = \sum_i^{\infty} |\psi_i(x)|^2 f_i, \ \ i = 1,2,...,\infty,
\label{rhodef}
\end{equation}
with occupation numbers $0 \leq f_i \leq 2$, $i= 1,2,...\infty$.
The occupation numbers in (\ref{rhodef}) can be chosen according to the 
Fermi-Dirac distribution function
\begin{equation}
f_i= f_{\beta} (\varepsilon_i - \mu) = \frac{2}{1+e^{\beta(\varepsilon_i-\mu)}},
\label{fermidirac}
\end{equation}
where $\mu$ is the chemical potential chosen to ensure that
\begin{equation}
  \int \hat{\rho}(x) dx = N_e,
\label{chargesum1}
\end{equation}
and $\beta$ is the inverse of the temperature, i.e.,
$\beta = 1/(k_B T)$ with $k_B$ being the Boltzmann constant.

Note that $\hat{\rho}(x)$ is simply the diagonal of the single particle density matrix
defined by
\begin{equation}
\hat{\gamma}(x,x')=\sum_{i=1}^{\infty}
\psi_{i}(x)f_{\beta}(\varepsilon_i-\mu) \psi_i^{\ast}(x') \label{gammaeq},
\end{equation}
and the charge sum rule in (\ref{chargesum1}) can be expressed alternatively
by
\begin{equation}
\mbox{Tr}\left[\hat{\gamma}(x,x')\right] = N_e,
\label{chargesum2}
\end{equation}
where Tr denotes the trace of an operator.

It follows from (\ref{kseqs}) and (\ref{gammaeq}) that the electron
density $\hat{\rho}(x)$
is a fixed point of the Kohn-Sham map defined by
\begin{equation}
  \hat{\rho}(x) = \diag \left( f_{\beta} (\hat{H}[\hat{\rho}(x)] - \mu \delta(x,x'))\right),
\label{ksmap}
\end{equation}
where $\mu$ is chosen to satisfy (\ref{chargesum2}).
The most widely used algorithm for finding the solution to
(\ref{chargesum2}) and (\ref{ksmap}) is a Broyden type of quasi-Newton
algorithm.
In the physics literature, this is often referred to as the self-consistent
field (SCF) iteration. The most time consuming part of this algorithm is
the evaluation of $\hat{\rho}(x) = \hat{\gamma}(x,x)$ in (\ref{ksmap}).

\subsection{Basis expansion by nonorthogonal basis functions}
An infinite-dimensional Kohn-Sham problem can be discretized in a number of
ways (e.g., planewave expansion, finite difference, finite element
etc.).
In this paper, we focus on a discretization scheme in which
a Kohn-Sham orbital $\psi_i$ is expanded by a linear
combination of a finite number of basis functions $\{\varphi_j\}$, i.e.,
\begin{equation}
  \psi_{i}(x)=\sum_{j=1}^{N} \varphi_{j}(x) c_{ji}.
  \label{eqn:psiphi}
\end{equation}
We should note that the total number of basis functions $N$ is generally proportional
to the number of electrons $N_e$ or atoms in the system to be studied.
These basis functions $\{\varphi_{j}\}$ can be constructed to have local nonzero
support. But they may not necessarily be orthonormal to each other.
Examples of these basis functions include Gaussian type
orbitals~\cite{FrischPopleBinkley1984,VandeVondeleKrackMohamedEtAl2005}
and local atomic
orbitals~\cite{Junquera:01,MohanChen2010,MohanChen2011,KennyHorsfieldFujitani2000,Ozaki:03,BlumGehrkeHankeEtAl2009},
adaptive curvilinear coordinates~\cite{TsuchidaTsukada1998}, optimized
nonorthogonal
orbitals~\cite{BowlerMiyazakiGillan2002,FattebertBernholc2000,HineHaynesMostofiEtAl2009}
and  adaptive local basis functions~\cite{LinLuYingE2012}.  In numerical
examples presented in section~\ref{sec:examples}, we use a
set of nonorthogonal local atomic orbitals.

Substituting (\ref{eqn:psiphi}) into (\ref{kseqs}) yields
a generalized eigenvalue problem
\begin{equation}
  H C = S C \Xi,
  \label{eqn:nonortho}
\end{equation}
where $C$ is an $N \times N$ matrix with $c_{ij}$ being its $(i,j)$th entry,
$\Xi$ is a diagonal matrix with $\varepsilon_{i}$ on its diagonal,
$S_{ij}=\braket{\varphi_{i}}{\varphi_{j}}$, and
$H_{ij}=\averageop{\varphi_{i}}{\hat{H}}{\varphi_{j}}$. For
orthogonal basis functions, the overlap matrix $S$ is an identity matrix,
and Eq.~\eqref{eqn:nonortho} reduces to a standard eigenvalue problem.
When local atomic orbitals are used as the basis, $S$ is generally not
an identity matrix, but both $H$ and $S$ are sparse.

Without loss of generality, we assume the basis
functions and the Kohn-Sham orbitals to be real in the following
discussion.  Let $\Psi=[\psi_1,\cdots,\psi_{N}]$ and
$\Phi=[\varphi_1,\cdots,\varphi_{N}]$, Then Eq.~\eqref{eqn:psiphi} can be
written in a compact form
\begin{equation}
  \Psi=\Phi C.
  \label{}
\end{equation}
Consequently,
the single particle density matrix (\ref{gammaeq})
becomes~\cite{FattebertBernholc2000}
\begin{equation}
  \begin{split}
		\hat{\gamma}(x,x')&= \Psi(x) f_{\beta}(\Xi-\mu) \Psi^T(x')\\
 & = \Phi(x) C
 f_{\beta}(\Xi-\mu) C^{T} \Phi^T(x').
  \end{split}
  \label{eqn:compactgamma}
\end{equation}

\subsection{Pole expansion and selected inversion for nonorthogonal
basis functions}\label{subsec:pole}
The most straightforward way to evaluate $\hat{\gamma}(x,x')$ is
to follow the right hand side of (\ref{eqn:compactgamma}), which
requires solving the generalized eigenvalue problem (\ref{eqn:nonortho}).
The computational complexity of this approach is $\Or(N^3)$. This approach
becomes prohibitively expensive when the number of electrons or atoms in the
system increases.

An alternative way to evaluate $\hat{\gamma}(x,x')$, which circumvents
the cubic scaling of the diagonalization process, is to approximate
$\hat{\gamma}(x,x')$ by a Fermi operator expansion (FOE)
method~\cite{Goedecker1993}. In an FOE scheme, the function
$f_{\beta}(\Xi-\mu)$ is approximated by a linear combination of a number of simpler
functions, each of which can be evaluated directly without
diagonalizing the matrix pencil $(H,S)$.  A variety of FOE schemes have been
developed. They include polynomial expansion~\cite{Goedecker1993}, rational
expansion~\cite{LinLuYingE2009,BaroniGiannozzi1992,Ozaki2007}, and a hybrid scheme in
which both polynomials and rational functions are used~\cite{parrinello2008,LinLuCarE2009}.  In all these schemes, the number of simple functions used
in the expansion is
asymptotically determined by $\beta\Delta E$, where $\Delta E=\max_{i=1}^{N}
\abs{\varepsilon_i-\mu}$ is the spectrum width for the discrete problem.
An upper bound of $\Delta E$ can be obtained inexpensively by a
very small number of Lanczos steps~\cite{Lanczos1950}.

\begin{figure}[h]
  \begin{center}
    \includegraphics[width=0.40\textwidth]{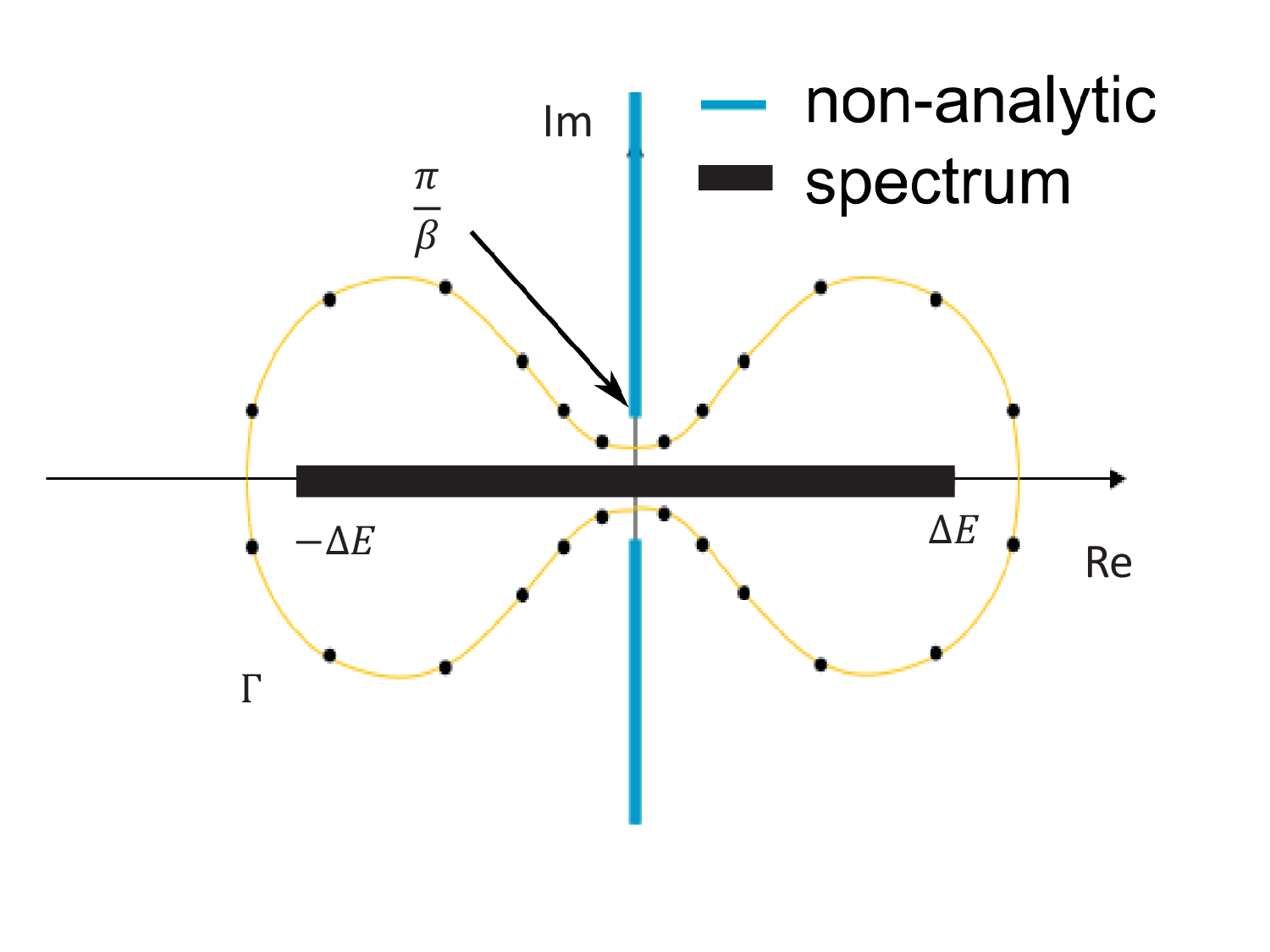}
  \end{center}
  \caption{(color online) A schematic view of the placement of poles
  used in a pole expansion approximation of $f_{\beta}(z)$.
  The thick black line on the real
  axis indicates the range of $\varepsilon_{i}-\mu$, and the thin blue
  line on the imaginary axis indicates the non-analytic part of
  $f_{\beta}(z)$.  The yellow dumbbell shaped contour is
  chosen to exclude the non-analytic part of the complex plane.
  Each block dot on the contour corresponds to a pole used in the
  pole expansion approximation.}
  \label{fig:pole}
\end{figure}

While most of the FOE schemes require as many as $\Or(\beta\Delta E)$ or
$\Or(\sqrt{\beta\Delta E})$ terms of simple functions, the recently
developed pole expansion~\cite{LinLuYingE2009} is particularly promising
since it requires only $\Or(\log \beta\Delta E)$ terms of simple rational
functions. 
The favorable scaling of the pole expansion allows us to treat
both insulating and metallic systems efficiently at room or
even lower temperature.
The pole expansion has the analytic expression 
\begin{equation}
  f_{\beta}(\varepsilon-\mu) \approx \Im \sum_{l=1}^{P}
  \frac{\omega^{\rho}_l}{\varepsilon-(z_l+\mu)}, \label{eqn:polerho}
\end{equation}
where 
\begin{equation}
	w_{l}^{\rho}=\frac{4K\sqrt{mM}}{\pi k P}
	\frac{\mathrm{cn}(t_l)\mathrm{dn}(t_l)}{z_{l}(k^{-1}-\mathrm{sn}(t_l))^2}
	f_{\beta}(z_l),
	\label{eqn:poleweight}
\end{equation}
with $m=\frac{\pi^2}{\beta^2}, M=\Delta E^2 + \frac{\pi^2}{\beta^2},
k=\frac{\sqrt{M/m}-1}{\sqrt{M/m}+1}$. The functions 
$\mathrm{cn},\mathrm{dn},\mathrm{sn}$ are Jacobi elliptic functions,
and $K,\{z_{l}\}, \{t_{l}\}$ are chosen carefully and computed from analytic
expressions. We refer the readers to
Ref.~\onlinecite{LinLuYingE2009} for more detailed explanations.  In the
following discussions, we will also refer
to $\{z_{l}\}$ as the complex shifts or \textit{poles}, and refer to
$\{\omega^{\rho}_l\}$ as the complex weights.  The complex shifts and
weights are determined only by $\beta,\Delta E$ and the number of poles
$P$. All quantities in the
pole expansion are known explicitly and their calculation takes
negligible amount of time.  The construction of pole expansion
is based on the observation that the non-analytic part of the
Fermi-Dirac function lies only on the imaginary axis within $\left[
\frac{i\pi}{\beta},+i\infty  \right]\bigcup \left[ -i\infty
,-\frac{i\pi}{\beta} \right]$. A dumbbell-shaped Cauchy contour (see
Fig.~\ref{fig:pole}) is carefully chosen and discretized to circle the
eigenvalues $\{\varepsilon_{i}\}$ on the real axis, while avoiding the
intersection with the non-analytic region. The pole expansion does not
require a band gap between the occupied and unoccupied states.
Therefore, it is applicable to both insulating and metallic systems.
Furthermore, the construction of the pole expansion relies only on the
analytical structure of the Fermi-Dirac function rather than its
detailed shape. This is a key property that is crucial for constructing
pole expansions for other functions, including the free energy density
matrix and the energy density matrix which are discussed in
section~\ref{subsec:energy} for the purpose of computing Helmholtz free
energy and atomic forces (including both the Hellman-Feynman force and
the Pulay force). In such case, one only needs to substitute $f_{\beta}$
in the weight function in Eq.~\eqref{eqn:poleweight} by the
corresponding function that shares the same analytic structure as the
Fermi-Dirac function $f_{\beta}$.


Following the derivation in the appendix, we can use (\ref{eqn:polerho}) to
approximate the single particle density matrix $\hat{\gamma}$ by its
$P$-term pole expansion, denoted by
$\hat{\gamma}_{P}$ as

\begin{equation}
  \begin{split}
    \hat{\gamma}_{P}(x,x') &= \Phi(x) \Im\left(
    \sum_{l=1}^{P}\frac{\omega^{\rho}_l}{H - (z_l+\mu) S}\right)
    \Phi^T(x')\\
    &\equiv \Phi(x) \Gamma \Phi^T(x').
  \end{split}
  \label{eqn:gammapole}
\end{equation}
In the above expression, $\Gamma$ is an $N\times N$ matrix represented
in terms of the atomic orbitals $\Phi$.  To simplify our notation,
we will drop the subscript $P$ from the $P$-term pole expansion
approximation of single particle density matrix $\hat{\gamma}$ unless
otherwise noted.  Similar treatment will be made for the
electron density $\hat{\rho}$, the total energy $E_{\tot}$, the
Helmholtz free energy $\mc{F}_{\tot}$, and the atomic force on the $I$-th
nuclei $F_{I}$.
Using Eq.~\eqref{eqn:gammapole}, we can evaluate the electron density in the
real space as the diagonal elements of
$\hat{\gamma}$, i.e.,
\begin{equation}
    \hat{\rho}(x) = \Phi(x) \Gamma \Phi^T(x)
     = \sum_{ij}\Gamma_{ij}\varphi_{j}(x)\varphi_{i}(x).
  \label{eqn:approxrho}
\end{equation}
We assume that each basis function $\varphi_{i}(x)$ is compactly
supported in the real space. In order to
evaluate $\hat{\rho}(x)$ for any particular $x$, we only need
$\Gamma_{ij}$ such that $\varphi_{j}(x)\varphi_{i}(x)\ne 0$, or
$S_{ij}\ne 0$. This set of $\Gamma_{ij}$'s is a subset of
$\{\Gamma_{ij}\vert H_{ij}\ne 0\}$. To obtain these {\em selected elements},
we need to compute the corresponding elements of
$(H - (z_l+\mu)S)^{-1}$ for all $z_l$.

The recently developed selected inversion
method~\cite{LinLuYingCarE:09,LinYangMezaEtAl2011,LinYangLuEtAl2011}
provides an efficient way of computing the selected elements of an
inverse matrix. For a symmetric matrix of the form $A=H-zS$,
the selected inversion algorithm
first constructs an $LDL^T$ factorization of $A$, where $L$ is a block
lower diagonal matrix called the Cholesky factor, and $D$ is a block
diagonal matrix.  In the second step, the selected inversion algorithm
computes all the elements $A^{-1}_{ij}$ such that $L_{ij}\ne 0$.  Since
$L_{ij}\ne 0$ implies that $H_{ij}\ne 0$, all the selected elements of
$A^{-1}$ required in \eqref{eqn:approxrho} are computed. As a result,
the computational scaling of the
selected inversion algorithm is only proportional to the number of
nonzero elements in the Cholesky factor $L$.  In particular, the
selected inversion algorithm has a complexity of $\Or(N)$ for quasi-1D systems,
$\Or(N^{1.5})$ for quasi-2D systems, and $\Or(N^{2})$ for 3D bulk
systems. The selected inversion algorithm achieves universal improvement
over the diagonalization method for systems of all dimensions.  It
should be noted that selected inversion algorithm is an \textit{exact}
method for computing selected elements of $A^{-1}$ if exact
arithmetic is to be employed, and in practice the only source of error
is the roundoff error.  In particular, the selected inversion algorithm
does not rely on any localization property of $A^{-1}$. However,
it can be combined with localization properties of insulating systems
to further reduce the computational cost.  We will pursue this approach
in future work.  We also remark that the PEXSI technique can be applied whenever $H$ and $S$ are sparse matrices.
However, since the selected inversion method relies on an $LDL^T$
factorization of $H-zS$, the preconstant of the selected inversion method
asymptotically scales cubically with respect to the number of basis
functions per atom.  The number of basis functions or degrees of freedom
per atom associated with the finite difference method~\cite{ChelikowskyTroullierSaad1994} and the finite element method~\cite{TsuchidaTsukada1995} is usually
much larger than that associated with methods based on contracted basis
functions such as local atomic orbitals.  Therefore the finite difference
method and the finite element method do not benefit as much from the PEXSI
technique as methods that are based on local atomic orbitals.

\subsection{Total energy, Helmholtz free energy and atomic force evaluation}
\label{subsec:energy}

In addition to reducing the computational complexity of the charge density
calculation in each SCF iteration, the PEXSI technique can also be used
to compute the total energy, the Helmholtz free energy as well as the atomic
forces (including both the Hellman-Feynman force and the Pulay
force) efficiently without diagonalizing the Kohn-Sham Hamiltonian.

It is well known that Eqs.~\eqref{kseqs}-~\eqref{chargesum1} can be 
derived as the first order necessary condition for minimizing the
Mermin free energy~\cite{Mermin1965,SolerArtachoGaleEtAl2002,KresseFurthmuller1996,WeinertDavenport1992,WentzcovitchMartinsAllen1992}
\begin{equation}
	\begin{split}
		\mc{F}_{\tot}\left[ \{\psi_{i}\}, \{f_i\} \right] =&   
		E_{\tot}\left[ \{\psi_{i}\}, \{f_i\} \right]  - T S\left[ \{f_i\}
		\right],
	\end{split}
	\label{eqn:ftotks}
\end{equation}
under the constraints~\eqref{orth} and $\sum_{i=1}^{\infty} f_i = N_{e}$,  
where
\begin{equation}
  \begin{split}
		E_{\tot}\left[ \{\psi_{i}\},\{f_i\} \right] =&
		\sum_{i=1}^{\infty}f_i \varepsilon_{i}
  - \frac12 \iint \frac{\hat{\rho}(x)
  \hat{\rho}(y)}{\abs{x-y}} \ud x \ud y \\
  &+ E_{\xc}[\hat{\rho}]
  - \int V_{\xc}[\hat{\rho}](x) \hat{\rho}(x) \ud x
  \end{split}
	\label{eqn:etotks}
\end{equation}
is called the internal energy or the total energy,
\begin{equation}
	S\left[ \{f_i\} \right] = 
	-2 k_B \sum_{i=1}^{\infty} \left( \tilde{f}_i \log \tilde{f}_i +
	(1-\tilde{f}_i) \log(1-\tilde{f}_i)\right)
	\label{eqn:entropy}
\end{equation}
is the entropy due to fractional occupation where $\tilde{f_i}=f_i/2$ is
used so that $0\le \tilde{f}_i\le 1$. The chemical potential $\mu$
in~\eqref{fermidirac} is simply the Lagrange multiplier associated with 
occupation number constraint $\sum_{i=1}^{\infty} f_i = N_{e}$.

Furthermore, it is the derivative of
the Mermin free energy (rather than the total energy) with respect to the
atomic positions that give rise to the correct force in ab initio
molecular dynamics simulation~\cite{SolerArtachoGaleEtAl2002,KresseFurthmuller1996,WeinertDavenport1992,WentzcovitchMartinsAllen1992}.  



The evaluation of the Mermin free energy functional $\mc{F}_{\tot}$ requires the
explicit knowledge of the Kohn-Sham eigenvalues $\{\varepsilon_{i}\}$
which are not available in the PEXSI scheme.  However, it has been shown in
Ref.~\onlinecite{AlaviKohanoffParrinelloEtAl1994} that the Mermin free
energy can be equivalently computed in the form of the following
Helmholtz free energy, which does not contain the Kohn-Sham eigenvalues
explicitly
\begin{equation}
  \begin{split}
  \mc{F}_{\tot} = &- 2\beta^{-1} \Tr \ln ( 1+ \exp(\beta( \mu -
  \Xi))) + \mu N_e  \\
  &- \frac12 \iint \frac{\hat{\rho}(x)
  \hat{\rho}(y)}{\abs{x-y}} \ud x \ud y + E_{\xc}[\hat{\rho}]\\
  &
  - \int V_{\xc}[\hat{\rho}](x) \hat{\rho}(x) \ud x.
  \end{split}
  \label{eqn:Helmholtz}
\end{equation}
Here we
assume LDA~\cite{CeperleyAlder1980} or
GGA~\cite{Becke1988,LeeYangParr1988} exchange-correlation functional is
used for the Kohn-Sham total energy expression. In
section~\ref{subsec:pole} we have shown that the electron density
$\hat{\rho}(x)$ can be computed in the PEXSI scheme.  
Therefore in Eq.~\eqref{eqn:Helmholtz}, only the first term requires
extra treatment. Note that the function
\begin{equation}
  f^{\mc{F}}_{\beta}(\varepsilon-\mu)=- 2\beta^{-1} \ln ( 1+ \exp(\beta(
  \mu -\varepsilon)))
  \label{}
\end{equation}
is different from the Fermi-Dirac function $f_{\beta}$ in Eq.~\eqref{fermidirac}.
In fact $f^{\mc{F}}_{\beta}$ is directly related to the $f_{\beta}$ as
\begin{equation}
	\left( f^{\mc{F}}_{\beta} \right)'(z) = f_{\beta}(z).	
	\label{}
\end{equation}
Nonetheless $f^{\mc{F}}_{\beta}(z)$ is analytic everywhere in the complex plane, except for segments of the
imaginary axis within $\left[ \frac{i\pi}{\beta},+i\infty  \right]\bigcup \left[
-i\infty ,-\frac{i\pi}{\beta} \right]$. In this sense,
$f^{\mc{F}}_{\beta}$ shares the same analytic structure as that of the
Fermi-Dirac function $f_{\beta}$. The pole expansion technique can be applied
with the same choice of poles $\{z_l\}$ but different weights, denoted by
$\{\omega_{l}^{\mc{F}}\}$, \ie
\begin{equation}
  f^{\mc{F}}_{\beta}(\varepsilon-\mu) \approx \Im \sum_{l=1}^{P}
  \frac{\omega^{\mc{F}}_l}{\varepsilon-(z_l+\mu)}.
  \label{eqn:poleHelmholtz}
\end{equation}
Following the derivation in the appendix, we can rewrite the Helmholtz free
energy as
\begin{equation}
  \begin{split}
  \mc{F}_{\tot} =&  \Tr[\Gamma^{\mc{F}} S] + \mu N_e
  - \frac12 \iint \frac{\hat{\rho}(x)
  \hat{\rho}(y)}{\abs{x-y}} \ud x \ud y\\
  &
  + E_{\xc}[\hat{\rho}]
  - \int V_{\xc}[\hat{\rho}] \hat{\rho}(x) \ud x,
  \end{split}
  \label{eqn:Helmholtzpole}
\end{equation}
where the free energy density matrix $\Gamma^{\mc{F}}$ is
given by
\begin{equation}
  \Gamma^{\mc{F}} = \Im \sum_{l=1}^{P}
  \frac{\omega^{\mc{F}}_l}{H - (z_l+\mu) S}.
  \label{eqn:gammaF}
\end{equation}
Note that in the expression~\eqref{eqn:Helmholtzpole}, the
first term depends on the trace of the product of $\Gamma^{\mc{F}}$ and
$S$. The computation of this term requires only the $(i,j)$th entry of
$\Gamma^{\mc{F}}$ for $(i,j)$ satisfying $S_{ij} \neq 0$ or
$H_{ij} \neq 0$.  Since
the poles $\{z_{l}\}$ are the same as those used for computing the
electron density, the selected elements of $\Gamma^{\mc{F}}$ correspond
to the same selected elements of $\left(H - (z_l+\mu) S\right)^{-1}$
used for the charge density calculation. Thus using them for computing
$\mc{F}_{\tot}$ does not introduce additional complexity.

It is worth mentioning that the above formulation can be simplified for
insulating systems with a relatively large band gap (even at zero temperature).  In such cases,
$f_{i}$ can be chosen to be $2$ for occupied states and $0$ for
unoccupied states.  Then the entropy term $S$ vanishes and
$\mc{F}_{\tot}=E_{\tot}$.  Furthermore, similar to the Helmholtz free
energy, an alternative expression for $E_{\tot}$ is
\begin{equation}
  \begin{split}
  E_{\tot} =&   \Tr[\Gamma H] - \frac12 \iint \frac{\hat{\rho}(x)
  \hat{\rho}(y)}{\abs{x-y}} \ud x \ud y \\
  &+ E_{\xc}[\hat{\rho}]
  - \int V_{\xc}[\hat{\rho}](x) \hat{\rho}(x) \ud x,
  \end{split}
  \label{eqn:energyformula}
\end{equation}
where $\Gamma$ is the density matrix defined in \eqref{gammaeq}.
Note that in this expression, the first term depends on the trace of
the product of $\Gamma$ and $H$. The computation of this term requires only
the $(i,j)$th entry of $\Gamma$ for $(i,j)$ satisfying $H_{ij} \neq 0$.
These entries are already available from the charge density calculation, thus
using them for total energy evaluation does not introduce additional
complexity.

To perform geometric optimization or ab initio molecular dynamics, we
need to compute atomic forces associated with different atoms.
Atomic force is the derivative of the free energy with respect
to the position of an atom.  For nonorthogonal atomic basis set, the
force calculation is not trivial, and standard methods have established
in Ref.~\onlinecite{SolerArtachoGaleEtAl2002} to calculate the force.
The calculation includes both the Hellman-Feynman force and the Pulay
force~\cite{Pulay:69}, where the Pulay force is induced by the change of
basis functions with respect to atomic positions.  Following the
derivation in the appendix, we can express the atomic force associated
with the $I$-th atom in a compact way as
\begin{equation}
	F_{I} = -\frac{\partial \mc{F}_{\tot}}{\partial R_I} =
  -\Tr\left[ \Gamma \frac{\partial H}{\partial R_I} \right]
      +\Tr\left[ \Gamma^E \frac{\partial S}{\partial R_I} \right].
  \label{eqn:forcepole}
\end{equation}
where $\Gamma^{E}$ is the energy density matrix defined by
\begin{equation}
  \Gamma^{E}=C \Xi f_{\beta}(\Xi-\mu)  C^T.
  \label{eqn:gammaE}
\end{equation}
We remark that Eq.~\eqref{eqn:forcepole} itself is not new. We
re-derive this formula in the appendix using linear algebra notation to
make the manuscript more accessible to readers not familiar with this
subject.  The concept of the energy density matrix has been used
before~\cite{SankeyNiklewski1989,SolerArtachoGaleEtAl2002}, and the last term in
Eq.~\eqref{eqn:forcepole} is also referred to as the ``orthogonalization
force'' in the appendix of Ref.~\onlinecite{SolerArtachoGaleEtAl2002},
which takes into account the fact that eigenfunctions must be
orthogonalized after atomic positions change.  

To illustrate more clearly that both the Hellman-Feynman force
and the Pulay force are taken into account correctly, let us look into the first term in
Eq.~\eqref{eqn:forcepole},
\begin{equation}
  \begin{split}
    \frac{\partial H_{ij}}{\partial R_{I}}&=\left\langle
    \frac{\partial\varphi_{i}}{\partial R_{I}},\hat{H}\varphi_{j}\right\rangle
    +\left\langle\varphi_{i},\frac{\partial{\hat{H}}}{\partial{R_{I}}}\varphi_{j}\right\rangle\\
    &+\left\langle\varphi_{i},\hat{H}
    \frac{\partial\varphi_{j}}{\partial R_{I}}\right\rangle.
  \end{split}
  \label{eqn:pulay}
\end{equation}
The terms $\frac{\partial\varphi_{i}}{\partial R_{I}}$ are automatically
included to reflect the change of the atom-centered basis functions with
respect to atomic positions, which gives rise to the Pulay force.
From a computational point of view, the terms in Eq.~\eqref{eqn:pulay} that
are related to the kinetic and non-local pseudopotential parts can be solved
by efficient two center integrals techniques,
while the terms related to local potential parts 
can be solved on a real space uniform grid.
The Hartree potential and the exchange correlation
potential are involved in the first term and the third term on the right
hand side of Eq.~\eqref{eqn:pulay}, but have no contribution to the
second term on the right hand side of Eq.~\eqref{eqn:pulay}.
Once all the terms in Eq.~\eqref{eqn:pulay} 
are evaluated,
one only needs to multiply them with density matrix $\Gamma$, 
which is obtained directly from the PEXSI method.

In order to compute the energy density matrix in Eq.~\eqref{eqn:gammaE}, 
and therefore the orthogonalization force without using the Kohn-Sham
eigenvalues $\{\varepsilon_{i}\}$ and Kohn-Sham
orbitals $\{\psi_i\}$, it is sufficient to note that the function
\begin{equation}
  f^{E}_{\beta}(\varepsilon-\mu) = \varepsilon f_{\beta}(\varepsilon-\mu)
  \label{eqn:fE}
\end{equation}
shares the same analytic structure as that of the Fermi-Dirac function
$f_{\beta}$. Thus, the energy density matrix can be approximated
by the same pole expansion used to approximate the density
matrix \eqref{eqn:gammapole}. In particular, there is no difference in the
choice of poles $z_l$.  But the weights of the expansion, which we denote by
$\omega_l^E$, for the energy density matrix approximation,
are different.
To be specific, the energy density matrix can be written using
the pole expansion as
\begin{equation}
  \Gamma^{E} = C \Im \sum_{l=1}^{P} \frac{\omega^{E}_l}{\Xi-(z_l+\mu) I}
  C ^T = \sum_{l=1}^{P} \frac{\omega^{E}_l}{H-(z_l+\mu)S}.
  \label{eqn:redEgamma}
\end{equation}
Again the selected elements of $\Gamma^E$ required in \eqref{eqn:forcepole}
can be easily computed from the selected elements of $[H-(z_l + \mu)S]^{-1}$
which are available from the charge density calculation.
\subsection{Chemical potential update}\label{subsec:mu}
The true chemical potential $\mu$ required in the pole expansions
\eqref{eqn:gammapole}, \eqref{eqn:Helmholtzpole} and \eqref{eqn:redEgamma}
is not known a priori. It must be solved iteratively as part of the solution
to \eqref{chargesum2} and \eqref{ksmap}.  For a fixed Hamiltonian $H$
associated with a fixed charge density, it is easy to show that
the left hand side \eqref{chargesum2}, which can be expressed as,
\begin{equation}
  N(\mu) = \Tr[\hat{\gamma}] = \Tr[\Gamma \Phi^T \Phi] = \Tr[\Gamma S]
  \label{eqn:Nmu}
\end{equation}
is a non-decreasing function with respect $\mu$. Hence the
root of \eqref{chargesum2} can be obtained by either Newton's method 
or the bisection method.  Other strategies for updating the
chemical potential have also been discussed in more detail in
literature~\cite{Goedecker1999,Ozaki2010}.

In an SCF iteration, $\hat{\rho}$ and $\mu$ are often updated in an
alternating fashion.  When the Kohn-Sham energies $\varepsilon_i$
associated with a fixed charge density are available, both
$N(\mu)$ and its derivative can be easily evaluated in Newton's method.
However, if $\hat{\gamma}$ is approximated via a pole expansion
\eqref{eqn:gammapole}, a new expansion is needed whenever $\mu$ is updated.
In Newton's method, the derivative of $N(\mu)$ can be approximated by
finite difference.  When $\mu^k$ is sufficiently close to the true chemical
potential, the derivative of $N(\mu^k)$ can be approximated by
\begin{equation}
  N'(\mu^k) \approx \frac{N(\mu^{k})-N(\mu^{k-1})}{\mu^{k}-\mu^{k-1}}.
	\label{eqn:newtonderivative}
\end{equation}
We remark that although
Newton's method converges rapidly near the correct chemical potential as
can be seen from the numerical results in section~\ref{sec:examples}, it
may not always be robust and may give very large correction when the
derivative~\eqref{eqn:newtonderivative} is small.  In such case 
a damped Newton's method or the
bisection method can be used instead to ensure the convergence of the
chemical potential iteration.  It remains challenging to update the
chemical potential both efficiently and robustly for all systems with
wide range of initial guesses, especially in the presence of gap states,
and dispersive bands which require global Fermi level finding across
multiple k-points.  We will develop efficient and robust schemes to
overcome this difficulty in our future work.

\subsection{Flowchart of PEXSI}

In Alg.~1 we summarize the main steps of the PEXSI technique
for accelerating atomic orbital-based electronic structure calculation
with the SCF iteration. We see that PEXSI replaces the diagonalization
procedure in solving KSDFT, and obtains the electron density, the total
energy, the Helmholtz free energy and the atomic force accurately
without computing eigenvalues and eigenfunctions of the Hamiltonian
operator.

\section{Numerical results} \label{sec:examples}
In this section, we report the performance achieved by applying
the PEXSI technique to an existing electronic structure calculation code
that uses local atomic orbital expansion to discretize the Kohn-Sham
equations.

\begin{figure}[h]
  \begin{center}
    \includegraphics[width=0.35\textwidth]{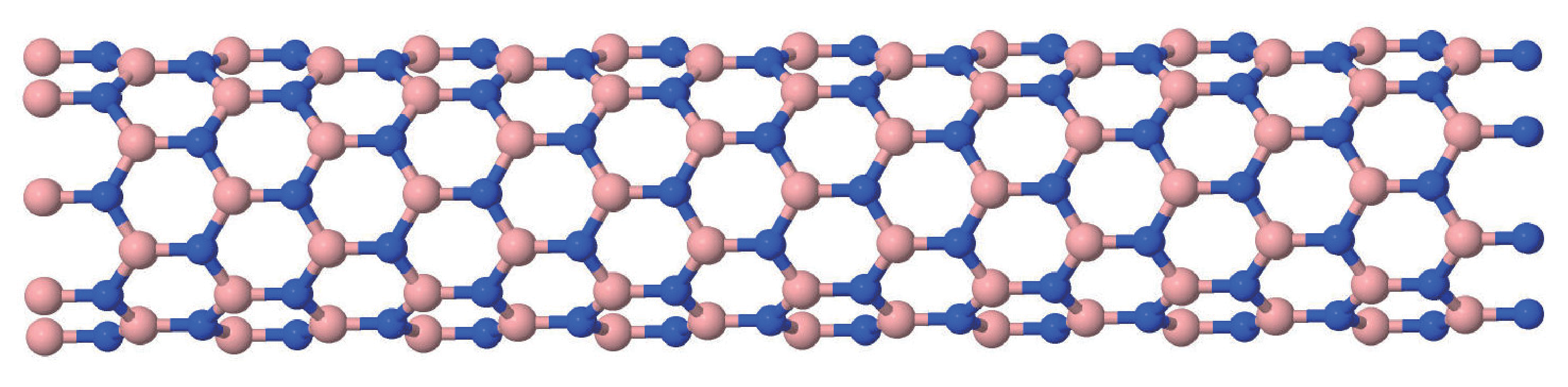}
  \end{center}
  \caption{(color online) Boron nitride nanotube (8,0) with 256 atoms. The
  boron atoms are labeled as pink (light gray) balls while the nitrogen
  atoms are labeled as blue (dark gray) balls. The bond length between a pair of
  adjacent boron and nitride atoms is 1.45 Angstrom.}
  \label{fig:BNNT256}
\end{figure}

\begin{figure}[h]
  \begin{center}
    \includegraphics[width=0.40\textwidth]{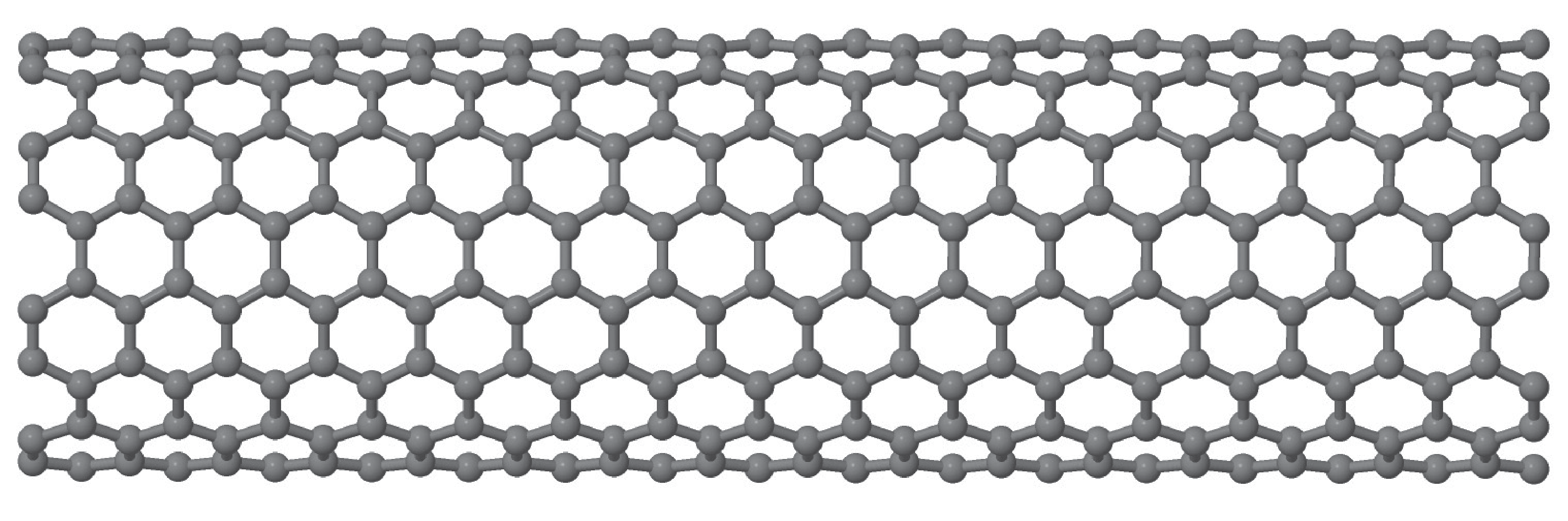}
  \end{center}
  \caption{(color online) Carbon nanotube (8,8) with 512 atoms. The carbon
  atoms are labeled as gray balls. The bond length between a pair of adjacent carbon
  atoms is 1.42 Angstrom.}
  \label{fig:CNT512}
\end{figure}

The test problems we used are two types of nanotubes.  One is a boron
nitride nanotube (BNNT) with chirality (8,0), which is an insulating system
shown in Figure~\ref{fig:BNNT256}.  The other is a carbon nanotube (CNT) with chirality
(8,8) shown in Figure~\ref{fig:CNT512}, which is a metallic system.
According to the formula $d=\frac{\sqrt{3}a}{\pi}\sqrt{n^{2}+mn+m^{2}}$,
where $a$ is the bond length and $(n,m)$ is the chirality of
nanotubes~\cite{CharlierBlaseRoche2007}, the diameter for BNNT (8,0) is
12.09 Bohr and for CNT (8,8) is 20.50 Bohr.
The longitudinal length of BNNT (8,0) with 256 atoms is roughly the same
as CNT (8,8) with 512 atoms.

	\begin{small}
  \begin{center}
    \begin{minipage}[t]{3.0in}
			\begin{tabular}{p{3.0in}}
				\hline
				\textbf{Algorithm 1}: Flowchart of the PEXSI technique.
			\end{tabular}
      \begin{tabular}{p{0.5in}p{2.5in}}
	{\bf Input}:  &  \begin{minipage}[t]{2.5in}
	  Atomic position $\{R_I\}$. Basis set $\Phi$. A subroutine to
	  construct matrices $H,S$ and matrices $\left\{\frac{\partial
	  H}{\partial R_{I}}\right\}, \left\{\frac{\partial S}{\partial
	  R_{I}}  \right\}$ given any electron density $\hat{\rho}$.
	\end{minipage} \\
	{\bf Output}:  &  \begin{minipage}[t]{2.5in}
	  Converged electron density $\hat{\rho}$. Total energy $E_{\tot}$.
	  Helmholtz free energy $\mc{F}_{\tot}$. Atomic forces
	  $\left\{ F_{I} \right\}$.  Chemical potential $\mu$.
	\end{minipage}
	\end{tabular}
    \begin{algorithmic}[1]
      \WHILE {$\hat{\rho}$ has not converged}
      \STATE Update $\hat{\rho}$ via charge mixing schemes for the SCF
			iteration.
      \STATE Construct matrices $H,S$ using the updated electron density
      $\hat{\rho}$.
      \WHILE {$\mu$ has not converged}
      \STATE Update the chemical potential $\mu$.
      \FOR {each pole $l=1,\ldots,P$}
      \STATE Compute the selected elements of each Green's function
      $\frac{1}{H-(z_{l}+\mu)}$ using selected inversion.
      \ENDFOR
      \STATE Compute $\Gamma$ via Eq.~\eqref{eqn:gammapole}, and
      compute the number of electrons $N(\mu)$ via Eq.~\eqref{eqn:Nmu}.
      \ENDWHILE
      \ENDWHILE
      \STATE Compute the free energy density matrix
      $\Gamma^{\mc{F}}$ via Eq.~\eqref{eqn:gammaF}, and the 
      energy density matrix $\Gamma^{E}$ via Eq.~\eqref{eqn:redEgamma}
      using the selected elements of the same set of Green's functions
      for computing $\Gamma$.
      \STATE Compute the converged electron density $\hat{\rho}$ via
      Eq.~\eqref{eqn:approxrho}, the total energy $E_{\tot}$ via
      Eq.~\eqref{eqn:energyformula}, the Helmholtz free energy $\mc{F}_{\tot}$ via
      Eq.~\eqref{eqn:Helmholtzpole}, and the atomic forces  $\left\{
      F_{I} \right\}$ via Eq.~\eqref{eqn:forcepole}.
    \end{algorithmic}
 \end{minipage}
  \begin{tabular}{p{3.0in}}
		\\
	\hline
	\end{tabular}
  \end{center}
\end{small}

We performed our
calculation at the Gamma point only.  Because Brillouin zone sampling can be
trivially parallelized, adding more $k$-points will not affect the
performance of our calculation.

Our computational experiments were performed on the Hopper system
at the National Energy Research Scientific Computing (NERSC) center.
The performance results reported below were obtained from running
the existing and modified codes on a single core of Hopper which
is part of a node that consists of two twelve-core AMD 'MagnyCours' 2.1-GHz
processors.  Each Hopper node has 32 gigabytes (GB) DDR3 1333-MHz memory. Each
core processor has 64 kilobytes (KB) L1 cache and 512KB L2 cache. It also
has access to a 6 megabytes (MB) of L3 cache shared among 6 cores.

Although the existing code has been parallelized using MPI and
ScaLAPACK, the parallelization of selected inversion is still
work in progress.  Hence, the performance study reported here is limited
to single processor runs.  However, we expect that the new approach of
using the PEXSI technique to compute
the charge density, total energy, Helmholtz free energy and force will
have a more favorable parallel scalability compared to diagonalizing the
Kohn-Sham Hamiltonian by ScaLAPACK because it can take advantage of an
additional level of parallelism introduced by the pole expansion. Due to
the availability of such parallelism, the cost of the computational time
of PEXSI is reported as the wall clock time for evaluating the selected
elements of one single pole.


In addition to comparing the performance of the existing and new approaches
in terms of wall clock time, we will also report the accuracy of our
calculation and memory usage.

\subsection{Atomic Orbitals and the Sparsity of $H$ and $S$}
The electronic structure calculation code we used for the performance study
is based on a local atomic orbital expansion
scheme~\cite{MohanChen2010,MohanChen2011}. We will refer to this scheme
as the CGH scheme below.  In the CGH scheme, an atomic orbital
$\varphi_{\mu}({\bf
r})$ is expressed as the product of a radial wave function $f_{\mu,l}(r)$
and a spherical harmonic $Y_{lm}(\hat{r})$,
where $\mu = \{\alpha, i, \zeta, l, m\}$, and $\alpha, i, \zeta, l, m$
represent the atom type, the index of an atom, the multiplicity
of the radial functions, the angular momentum and the magnetic quantum number
respectively.  The radial function $f_{\mu,l}(r)$ is constructed as a linear
combination
of spherical Bessel functions within a cutoff radius $r_{c}$, i.e.,
\begin{equation}
f_{\mu,l}(r)=\left\{
\begin{array}{ll}
\sum_{q}c_{\mu q}j_l(qr), & r < r_c\\
0 & r \geq r_c \, .\\
\end{array}
\right.
\end{equation}
where $j_l(qr)$ is a spherical Bessel function with $q$ chosen to satisfy
$j_{l}(qr_c)$=0, and the coefficients $c_{\mu q}j_l(qr)$ are chosen
to minimize a ``spillage
factor''~\cite{Sanchez-PortalArtachoSoler1995,Sanchez-PortalArtachoSoler1996}
associated with a reference system that consists of a set of (4 or 5)
dimers.  We refer readers to
Ref.~\onlinecite{MohanChen2010,MohanChen2011} for the details on the construction of the
CGH local atomic orbitals.

The cutoff radius $r_c$ determines the sparsity of the 
Kohn-Sham Hamiltonian $H$ and the overlap matrix $S$. The smaller the radius,
the sparser $H$ and $S$ are.  The cutoff radius for the atomic orbitals is set
to $8.0$ Bohr for B and N atoms in BNNT, and $6.0$ Bohr for C atoms
in CNT, respectively. 
The reasons why we choose a larger cutoff radius for B, N
atoms is that the spillage factor for the B and N atoms is larger than
that for the C atoms if $6.0$ Bohr cutoff is used for all atoms, which 
affects the accuracy of atomic orbitals.  In general, the cutoff radius
of most atomic orbitals can be chosen below $10$ Bohr.


Another parameter that affects the dimension of $H$ and $S$ is the
multiplicity $\zeta$ of the radial function $f_{\mu,l}(r)$. The
multiplicity determines the number of basis functions per atom. A higher
multiplicity results in larger number of basis functions per atom,
which in turn results in more rows and columns in $H$ and $S$. In our
experiments, we used both single-$\zeta$ (SZ) orbitals and
double-$\zeta$ plus polar orbitals (DZP). The number of local atomic
orbitals is 4 for SZ and 13 for DZP.

\begin{figure}[h]
  \begin{center}
    \subfloat[]{\includegraphics[width=0.22\textwidth]{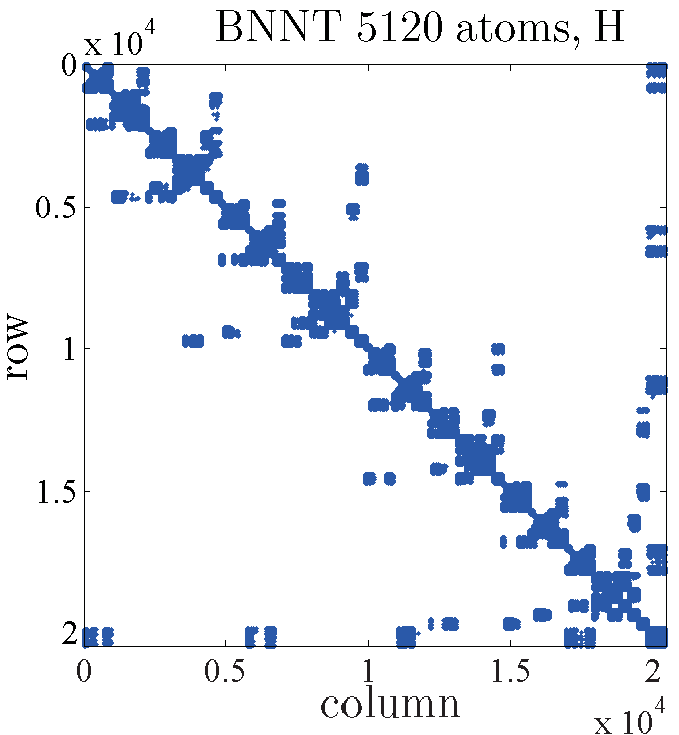}}\quad
    \subfloat[]{\includegraphics[width=0.22\textwidth]{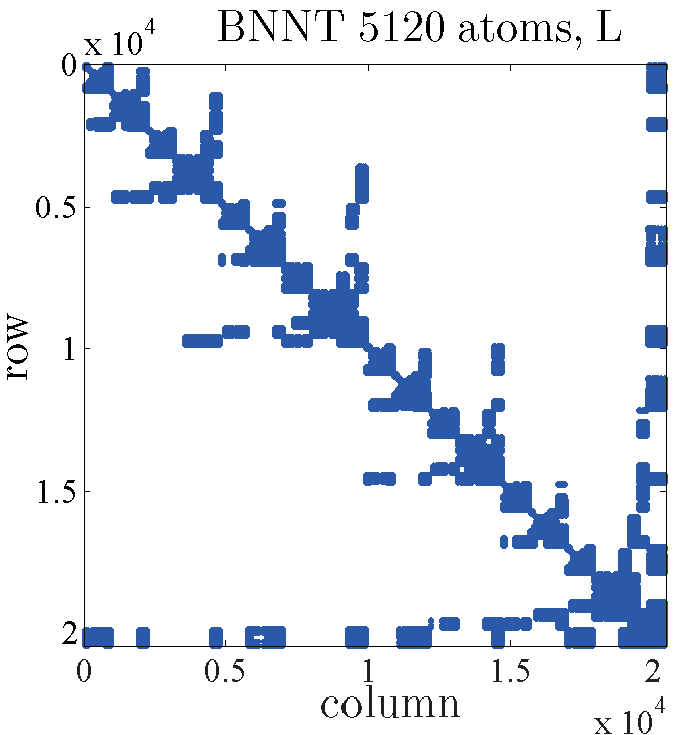}}
  \end{center}
  \caption{(color online) The sparsity pattern of $H$ (a) and $L+L^T$ (b)
  for an 5120-atom BNNT (8,0) with SZ orbitals.  Nested dissection reordering is used.}
  \label{fig:sparseBN}
\end{figure}

We measure the sparsity by the percentage of the nonzero elements in the
matrix $H$ denoted by
\begin{equation}
  \Hnnz = \frac{\mathrm{nnz}(H)}{N^2(H)} \times 100.
  \label{eqn:Hnnz}
\end{equation}
Here $\mathrm{nnz}(H)$ is the number of nonzero elements of $H$ and
$N(H)$ is the dimension of $H$ respectively.  Since the computational
cost of the selected inversion method is determined by
the sparsity of $L+L^T$ for the Cholesky factor $L$ of $H - z S$, we
will also report the percentage of the nonzero elements in the matrix $L+L^T$
(denoted by $\Lnnz$) below.  To reduce the amount of non-zero
fill-in of $L$, we use the nested dissection (ND) technique~\cite{George1973} to
reorder the sparse matrix $H-z S$ before it is factored.
Fig.~\ref{fig:sparseBN} (a) depicts the sparsity pattern of the $H$ matrix
associated with a 5120-atom BNNT (8,0) obtained from SZ atomic orbitals
after it is reordered by ND.  The sparsity pattern of $L+L^T$
for the corresponding Cholesky factor $L$ of the same problem
is shown in Fig.~\ref{fig:sparseBN} (b).

Table~\ref{tab:sparsity} shows the sparsity of Hamiltonian matrices
associated with BNNT (8,0) and CNT (8,8) systems that consist of $64$
to $10240$ atoms. The Hamiltonians for these systems are constructed from
SZ atomic orbitals.  We report both the $\Hnnz$ and $\Lnnz$
values. We can clearly see from this table that $H$, and consequently $L$,
are quite dense when the number of atoms in the
nanotubes is relatively small (less than 512). This is due to fact that
a large percentage of atoms in these small systems are within the $r_c$ distance
from each other.  When the system size becomes larger (with more than
$512$ atoms), both $\Hnnz$ and $\Lnnz$ are inversely proportional to
the system size.  This is because for quasi-1D systems, the numerator
in Eq.~\eqref{eqn:Hnnz} scales linearly with respect to $N(H)$ for large
$N(H)$.  Hence, the resulting matrices become increasingly
sparse, thereby making the selected inversion method more favorable.
\begin{table*}[htbp]
  \centering
  \begin{tabular}{c|c|c|c|c|c|c|c|c|c}
    \hline
    \multicolumn{2}{c|}{\# Atoms}  & 64 & 128 & 256 & 512 & 1024 & 1920 & 5120 & 10240 \\
    \hline
    \multirow{2}{*}{BNNT (8,0)} & $\Hnnz$ & 100.00 & 85.54 & 42.77 & 21.43 & 11.69 & 5.70 & 2.13 & 1.06\\
    & $\Lnnz$ & 100.00 & 99.48 & 77.94 & 46.13 & 25.07 & 13.70 & 5.26 & 2.64\\
    \hline
    \multirow{2}{*}{CNT (8,8)} & $\Hnnz$ & 40.63 & 38.67 & 19.53 & 9.77 & 4.88 & 2.60 & 0.97 & 0.49\\
    & $\Lnnz$ & 69.92 & 68.45 & 68.70 & 54.38 & 31.75 & 17.54 &
    7.42 & 3.79\\
    \hline
  \end{tabular}
  \caption{The percentage of nonzero elements $\Hnnz$ and $\Lnnz$ for
  BNNT (8,0) and CNT (8,8) of various sizes. }
  \label{tab:sparsity}
\end{table*}
%
%
%
%
%
\subsection{Performance comparison between diagonalization and selected inversion}
We now compare the efficiency of selected inversion with that of
diagonalization for computing the charge density in a single SCF iteration.
In the existing code, the diagonalization of the matrix pencil $(H,S)$
is performed by using the LAPACK subroutine {\tt dsygv} when the code
is run on a single processor.  The selected inversion is performed by
the {\tt SelInv} software~\cite{LinYangMezaEtAl2011}.


We use BNNT(8,0) and CNT(8,8) nanotubes of different lengths to study
the scalability of the computation with respect to the number of atoms
in the nanotube.  The number of atoms in these tubes ranges from $64$
to $10240$.

\begin{figure}[h]
  \begin{center}
    \includegraphics[width=0.45\textwidth]{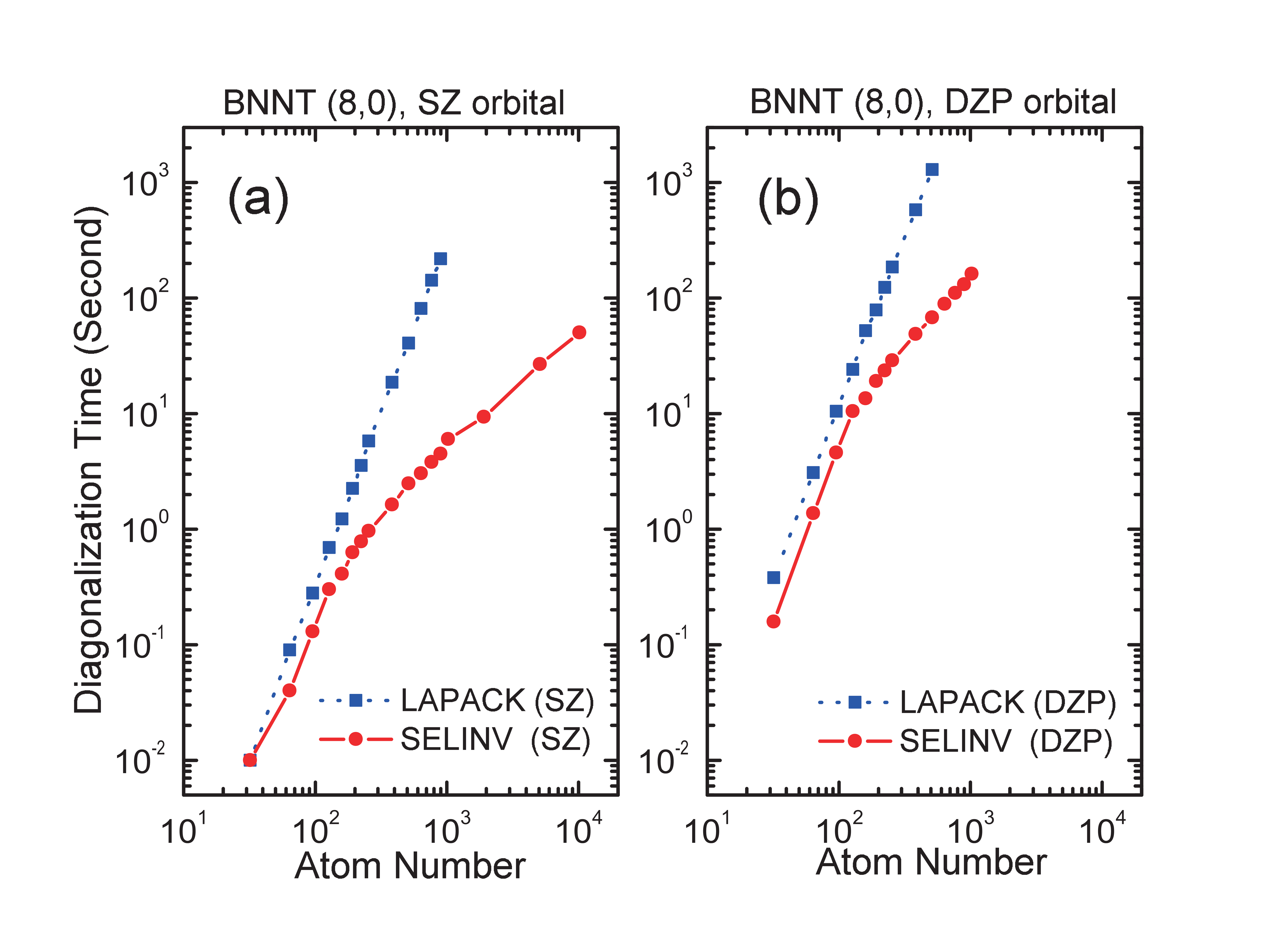}
  \end{center}
  \caption{(color online) Comparisons of the wall clock time used
  by selected inversion (at one pole) required for PEXSI and
  by the LAPACK {\tt dsygv} used to diagonalize a Kohn-Sham Hamiltonian
  associated with BNNT (8,0).  The Hamiltonians are
  constructed from SZ orbitals (4 basis per atom) in (a) and
  DZP orbitals (13 basis per atom) in (b).}
  \label{fig:CrossOverBN}
\end{figure}

Fig.~\ref{fig:CrossOverBN} shows how the wall clock time used by {\tt SelInv}
compares with that used by {\tt dsygv} for BNNT(8,0)
of different sizes.  When SZ atomic orbitals are used, {\tt SelInv}
takes
almost the same amount of time as that used by {\tt dsygv} for a BNNT
with $64$ atoms.
When the number of atoms is larger than 64, {\tt SelInv} is
more efficient than {\tt dsygv}.  The cubic scaling of {\tt dsygv} with
respect to the number of atoms can be clearly seen from the slope of the
blue loglog curve, which is approximately 3.  The linear scaling of
{\tt SelInv}, which is indicated by the slope of the red curve, is evident
when the number of atoms exceeds 200.  For systems with less than 200 atoms,
the wall clock time consumed by {\tt SelInv} scales cubically with
respect to the number of atoms also. This is due to the fact that the
$H$ and $S$ matrices associated with these small systems are nearly dense.
Similar observations can be made when the DZP atomic orbitals are used.
In this case, {\tt SelInv} is already more efficient than {\tt dsygv} when
the number of atoms is only 64.  The linear scaling of {\tt SelInv} can be
observed when the number of atoms exceeds $128$.

\begin{figure}[h]
  \begin{center}
    \includegraphics[width=0.45\textwidth]{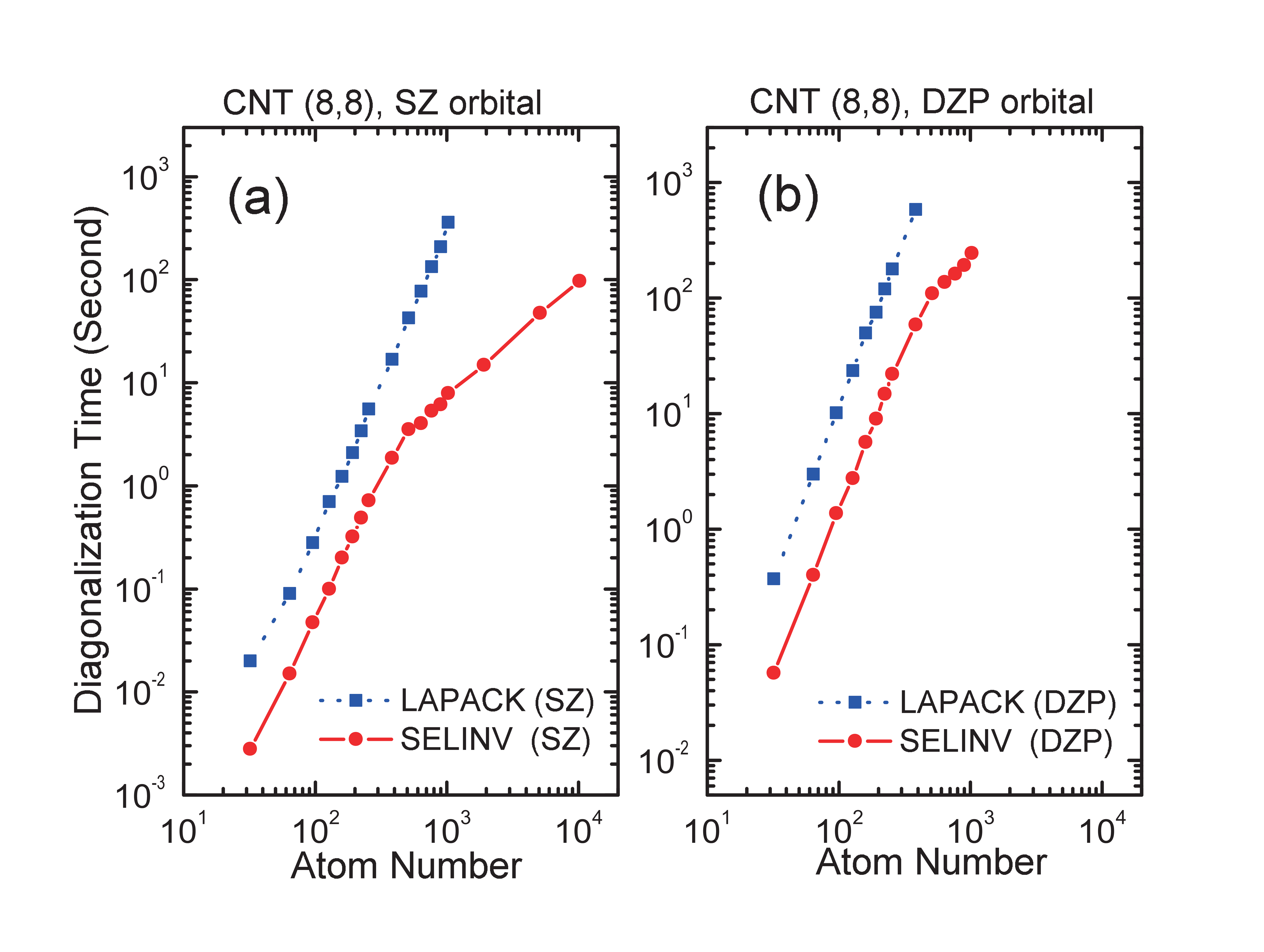}
  \end{center}
  \caption{(color online) Comparisons of the wall clock time
  by selected inversion (at one pole) required for PEXSI and
  by the LAPACK {\tt dsygv} used to diagonalize a Kohn-Sham Hamiltonian
  associated with CNT(8,8).  The Hamiltonians are
  constructed from SZ orbitals (4 basis per atom) in (a) and
  DZP orbitals (13 basis per atom) in (b).}
  \label{fig:CrossOverC}
\end{figure}

Fig.~\ref{fig:CrossOverC} shows the timing comparison between
{\tt SelInv} and {\tt dsygv} for CNT (8,8) of different sizes.
Because the cutoff radius for the carbon atom is chosen to be 6.0,
which is smaller than that associated with the boron and nitrogen atoms,
the $H$ and $S$ matrices associated with CNT (8,0) are sparser even when
the number of atoms in the tube is relatively small.  This explains
why {\tt SelInv} is already more efficient than {\tt dsygv} for
a CNT with $64$ atoms regardless whether SZ or DZP atomic orbitals are used.
However, the linear scaling of {\tt SelInv} timing with respect to
the number of atoms does not show up until the number of atoms reaches
500. The increase in the crossover point is due to the fact that the
sparsity of $H$ is asymptotically determined by the number of
atoms per unit length of the nanotube. Because the CNT (8,0) we use
in our experiment has a large diameter, there are more atoms along the
radial direction per unit length in CNT than that in BNNT.  Consequently,
it takes almost twice as many as atoms for CNT to reach the same length
along the longitudinal direction when compared to BNNT, as we can see
from Fig.~\ref{fig:BNNT256} and Fig.~\ref{fig:CNT512}.

We should note here that it is possible to combine the PEXSI
technique with a SZ atomic orbital based Kohn-Sham
DFT solver to perform electron structure calculation on quasi-1D
systems with more than 10,000 atoms. On the Hopper machine, the
wall clock time used to perform a single selected inversion of the $H-z S$
matrix associated with a 5,120-atom BNNT(8,0) is 26.72 seconds.
When the number of atoms increases to 10240, the wall clock time increases
to 50.07 seconds.  Similar performance is observed for CNT(8,8).
It takes 47.59 seconds to perform a selected inversion for a 5120-atom
CNT(8,8) tube, and 97.16 seconds for a 10240-atom tube.

\subsection{Memory usage}
We should also remark that the memory requirement for {\tt SelInv}
increases linearly with respect to the number of atoms when
the nanotube reaches a certain size. For a nanotube that consists of $10240$
atoms, the amount of memory required to store $L$ and the selected elements
of $[H-(z_{l}+\mu)S]^{-1}$ is $0.66$ GB and $0.93$ GB respectively.
The relatively
low memory requirement of {\tt SelInv} for quasi-1D system suggests that
the method may even be applicable to quasi-1D systems that contain more
than $100,000$ atoms on a single processor.

\subsection{Accuracy}
When selected inversion can be computed to high accuracy, which is often
the case in practice, the only source of error introduced by
the PEXSI technique comes from the limited number of terms in the pole
expansion \eqref{eqn:gammapole}.  The number of poles needed
in \eqref{eqn:gammapole} to achieve a desired level of accuracy in
total energy (or free energy) and force is largely determined
by the inverse temperature $\beta=1/(k_B T)$ used in~\eqref{fermidirac}
and the spectrum width $\Delta E$.  Here we show that at room
temperature $T = 300K$,
the number of poles required to provide an accurate pole expansion
approximation is modest even for a metallic system such as CNT(8,8).
Table~\ref{tab:poleaccuracy} shows that when diagonalization is replaced
by PEXSI for a single $\Gamma$-point calculation, the errors in total energy
and force decrease as the number of poles in \eqref{eqn:gammapole} increases.
The force difference is measured between the force
calculated by the PEXSI scheme using Eq.~\eqref{eqn:forcepole}, and that
calculated by the LAPACK diagonalization subroutine {\tt dsygv} using
standard methods~\cite{SolerArtachoGaleEtAl2002} previously implemented
in the CGH atomic orbital scheme~\cite{MohanChen2010,MohanChen2011}.
When the number of poles reaches 80, the difference between the final
total energies produced by the existing code and the modified code
(which replaces diagonalization with PEXSI) is $3.6\times 10^{-7}$ eV.
The difference in the mean absolute error (MAE) is $2\times 10^{-6}$
eV/Angstrom, which is quite small for all practical purposes.
\begin{table}[htbp]
  \centering
  \begin{tabular}{c|c|c}
    \hline
    \# Poles  & $E_{\textrm{PEXSI}}-E_{\textrm{ref}}$ (eV) & MAE Force (eV/Angstrom) \\
    \hline
    20 & 5.868351108 & 0.400431 \\
    40 & 0.007370583 & 0.001142 \\
    60 & 0.000110382 & 0.000026 \\
    80 & 0.000000360 & 0.000002 \\
    \hline
  \end{tabular}
    \caption{The difference between the total energy and atomic force produced
             by the existing electronic structure code and modified version
             in which diagonalization is replaced by PEXSI.
             The difference in atomic force is measured in terms of the mean
             absolute error (MAE).}
  \label{tab:poleaccuracy}
\end{table}

\begin{figure}[h]
  \begin{center}
    \includegraphics[width=0.45\textwidth]{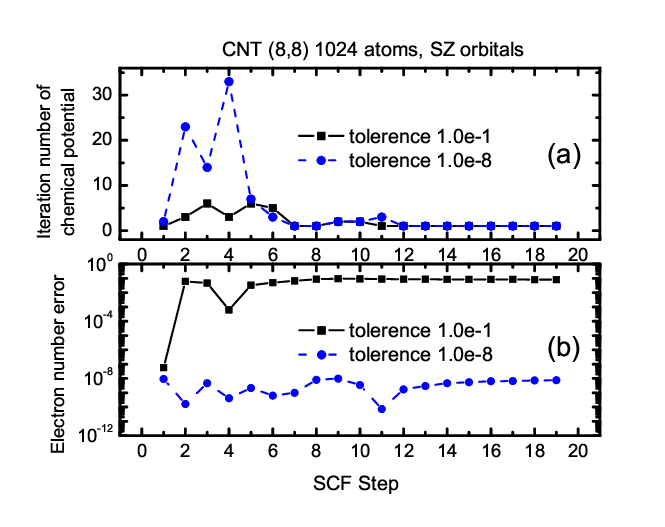}
  \end{center}
	\caption{(color online) The numbers of chemical potential
	iteration steps (a), and the error associated with the number of electrons (b) at
	different SCF iterations for CNT(8,8) with $1024$ atoms using SZ basis
	set.  The chemical potential is relaxed until the error of total
	number of electrons ($4096$ electrons
	in this system) is within $10^{-1}$ (blue dashed lines with dots) and
	within $10^{-8}$ (black solid lines with squares).}
  \label{fig:CNTMu}
\end{figure}

The numbers of chemical potential iterations, as well as the
error of the number of electrons at different SCF steps for a metallic
CNT(8,8) system with $1024$ atoms using SZ basis set is reported in
Fig.~\ref{fig:CNTMu}. The chemical potential is relaxed until the error
associated with the total electron number ($4096$ electrons in this system) is within a
given tolerance $\tau$. The average number of chemical potential
iterations is $2.01$ for the low accuracy case ($\tau=10^{-1}$), and
$5.21$ for the high accuracy case ($\tau=10^{-8}$), respectively.
Notice that in both cases, the number of chemical potential iterations
is $1\sim 2$ when the SCF gets close to convergence. Similar behavior is
also observed in the geometry optimization example in
section~\ref{subsec:geometry} for which the change of chemical potential
in consecutive steps is small. We further remark that the chemical
potential does not need to be performed very accurately at the first few
SCF steps. So the tolerance $\tau$ can be chosen dynamically with
respect to the accuracy of the current SCF step, in order to further
reduce the number of chemical potential iterations in the case of high
accuracy calculation.  We note that {\tt SelInv} is a direct
method for computing selected elements of the Green's function accurately.  
When low accuracy is allowed, it is possible to reduce the computational 
cost of this method further by discarding elements in the Cholesky factor 
with small magnitude.  This approach will be pursued in our future work.

%

\subsection{Overall Performance}
\begin{figure}[h]
  \begin{center}
    \includegraphics[width=0.45\textwidth]{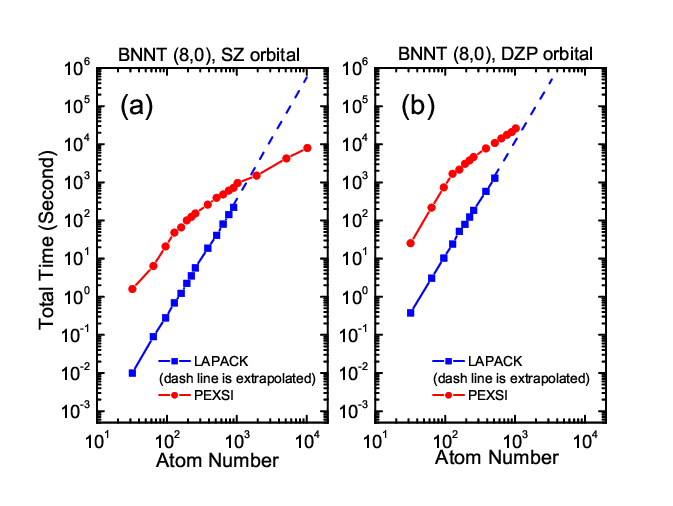}
  \end{center}
  \caption{(color online) Comparisons of the total wall clock time used
	to perform a PEXSI-based SCF iteration (using 80 poles and $2$
	iterations of chemical potential) and to perform an LAPACK {\tt dsygv}
	diagonalization based SCF iteration for BNNT (8,0) configured with
	different numbers of atoms.  The Hamiltonians are constructed from SZ
	orbitals (4 basis per atom) in (a) and DZP orbitals (13 basis per
	atom) in (b).}
  \label{fig:CrossOverBNTotal}
\end{figure}

\begin{figure}[h]
  \begin{center}
    \includegraphics[width=0.45\textwidth]{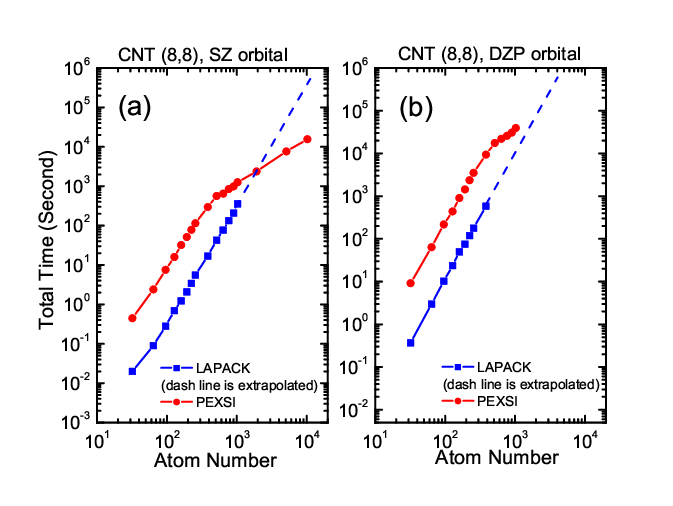}
  \end{center}
  \caption{(color online) Comparisons of the total wall clock time
  used to perform a PEXSI-based SCF iteration (using 80 poles and $2$ iterations of chemical 
  potential)  and an LAPACK {\tt dsygv} diagonalization
  based SCF iteration for CNT(8,8).  The Hamiltonians are
  constructed from SZ orbitals (4 basis per atom) in (a) and
	DZP orbitals (13 basis per atom) in (b).}
  \label{fig:CrossOverCTotal}
\end{figure}

One a sequential machine, the total wall clock time consumed by each
PEXSI-based SCF iteration is $t_{selinv} \times P \times k_{\mu}$, where
$t_{selinv}$ is the time required to perform one selected inversion, $P$
is the number of poles used in the pole expansion \eqref{eqn:polerho}
and $k_{\mu}$ is the average number of chemical potential iterations.
In practice, $P=80$ is often more than sufficient to yield an accurate
approximation in~\eqref{eqn:polerho} as we can see from
Table~\ref{tab:poleaccuracy}. The average $k_{\mu}$ can be $1\sim 2$
especially in geometry optimization and molecular dynamics.  If we take
$P=80$ and $k_{\mu} = 2$, the total wall clock time of a PEXSI-based
SCF iteration is compared with an LAPACK diagonalization based SCF iteration
for BNNT and CNT of various sizes in
Fig.~\ref{fig:CrossOverBNTotal} and~\ref{fig:CrossOverCTotal},
respectively. Since the LAPACK diagonalization routine cannot perform as
large of a calculation as PEXSI due to memory constraint, we extrapolate the
wall clock time of the LAPACK diagonalization routine in
Figures~\ref{fig:CrossOverBNTotal} and ~\ref{fig:CrossOverCTotal}, and we find that
the number of atoms beyond which the
sequential PEXSI method outperform the diagonalization method
is $1650$ atoms for BNNT(8,0) discretized  by SZ orbital, and $1800$
atoms for BNNT(8,0) discretized by DZP orbital. Similarly, the crossover
for the sequential PEXSI method to outperform the diagonalization method
is $1750$ atoms for CNT(8,8) discretized by SZ orbitals, and $1700$
atoms for CNT(8,8) discretized by DZP orbitals.

However, when a large number of processors are available, the
advantage of PEXSI becomes apparent.  Because each term in
\eqref{eqn:polerho} can be evaluated independently, we achieve
an automatic $P$-fold speedup whereas the speedup that can be
achieved by a parallel diagonalization procedure implemented in, for
example, the ScaLAPACK software package, is often limited.
Furthermore, each selected inversion can be parallelized, and our
current work, which we will publish in a separate publication, indicates
that excellent speedup can be achieved for this calculation on
hundreds of processors.  As a result, the PEXSI-based SCF iteration can
easily scale to tens of thousands of processors, whereas it is difficult
to make ScaLAPACK diagonalization procedures work efficiently on that
many processors.

%
%

\subsection{Geometry Optimization}\label{subsec:geometry}

\begin{figure}[h]
  \begin{center}
    \includegraphics[width=0.50\textwidth]{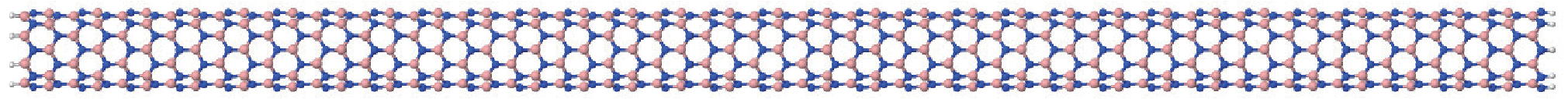}
  \end{center}
  \caption{(color online) A truncated boron-nitride nanotube (8,0) with
  1024 atoms, among which 504 boron atoms are labeled as pink (light
  gray) balls, 504 nitride atoms are labeled as blue (dark gray) balls,
  and 16 hydrogen atoms are labeled as small white balls.  The hydrogen
  atoms are used to passivate both ends of the nanotube.}
  \label{fig:1024tube}
\end{figure}

The PEXSI scheme with atomic orbitals can also be used for accurate
geometry relaxation of large-scale atomic systems.  We use a truncated
boron-nitride nanotube (8,0) with 1024 atoms, shown in
Fig.~\ref{fig:1024tube}, as an example to illustrate the efficiency of
PEXSI in this type of calculation. The nanotube contains 504 boron
atoms (B) and 504 nitride atoms (N). Each end of the nanotube is passivated
by $8$ hydrogen atoms (H).  We used DZP orbitals for all three atomic
elements.  The cutoff radius for B and N is set to 8.0 Bohr. The cutoff
radius for H is set to 6.0 Bohr.  We used 96 poles in the pole expansion
for both energy and force calculations.

\begin{figure}[h]
  \begin{center}
    \includegraphics[width=0.45\textwidth]{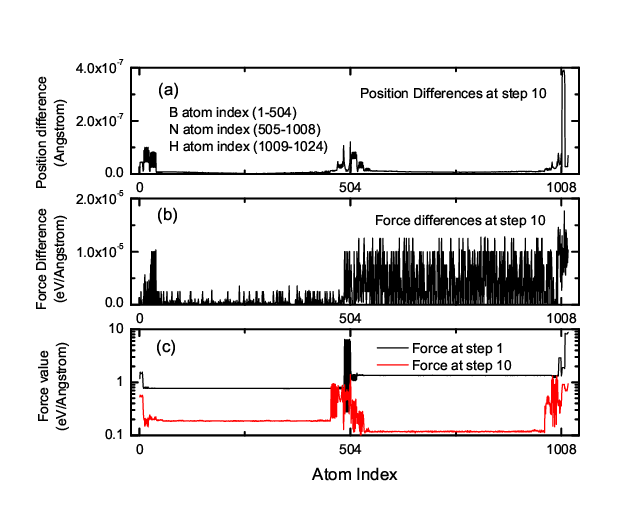}
  \end{center}
	\caption{(color online) The differences of the atomic
	positions (a) and forces (b) obtained from separate simulations
	using the PEXSI method and the diagonalization method, starting from
	the same initial condition. The result is obtained at the $10$-th
	geometry optimization step for the boron-nitride nanotube (8,0) system
	with 1024 atoms.  The absolute values of the forces at the $1$-st and
	the $10$-th geometry optimization steps are also presented (c). The
	tolerance for the error of the total number of electrons is chosen to
	be $10^{-8}$.}
  \label{fig:ForceDiff}
\end{figure}

Convergence is reached after 105 steps of ionic relaxation steps are
taken in the BFGS method.  The maximum atomic force associated with the
converged structure is less than 0.04 eV/Angstrom. 
To demonstrate the accuracy of the PEXSI method, we compare the
differences of the atomic positions and forces obtained from separate
geometry optimization simulations using the PEXSI method and the
diagonalization method, starting from the same initial condition.
Fig.~\ref{fig:ForceDiff} shows that 
at the $10$-th geometry optimization step, the maximum difference of the
atomic positions among all $1024$ atoms is less than $5\times
10^{-7}$ Angstrom (Fig.~\ref{fig:ForceDiff} (a)), and the maximum
difference of the forces is less than $2\times 10^{-5}$ eV/Angstrom
(Fig.~\ref{fig:ForceDiff} (b)).  Fig.~\ref{fig:ForceDiff} (c) shows
that at the $10$-th geometry optimization step the absolute value of the
force is still as large as $0.1\sim 1$ eV/Angstrom, and the relative
error of the forces obtained from the PEXSI method is around $0.01\%$.
This result shows that the PEXSI scheme is accurate for evaluating
the forces for this system.

\begin{figure}[h]
  \begin{center}
    \includegraphics[width=0.45\textwidth]{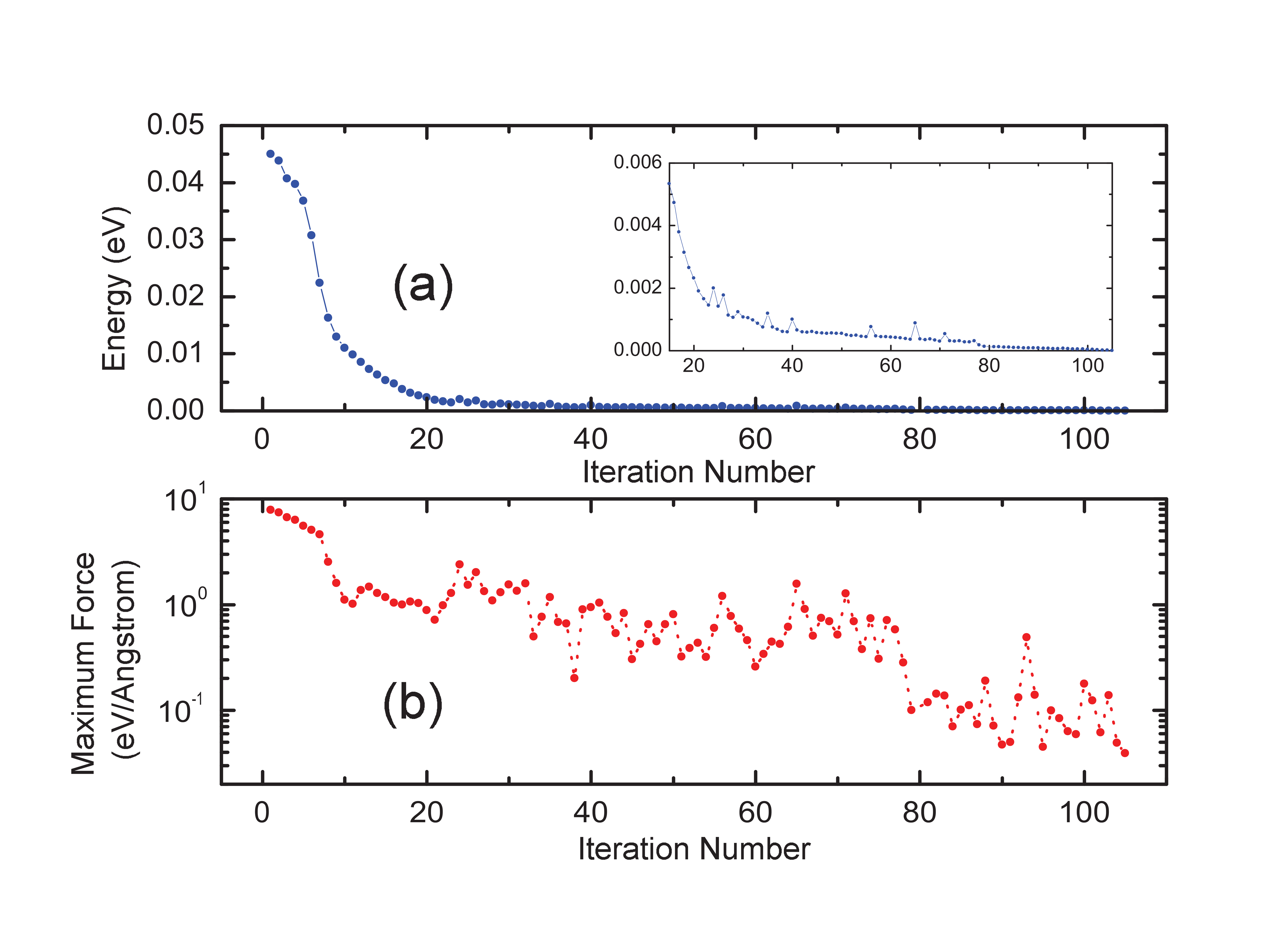}
  \end{center}
  \caption{(color online) The energy per atom (a) and the maximum force
  (b) for each geometry optimization iteration step. The criterion for
  the convergence of the force is set to 0.04 eV/Angstrom. The energy
  per atom at the last iteration step is set to zero.}
  \label{fig:EnergyForce}
\end{figure}

\begin{figure}[h]
  \begin{center}
    \includegraphics[width=0.45\textwidth]{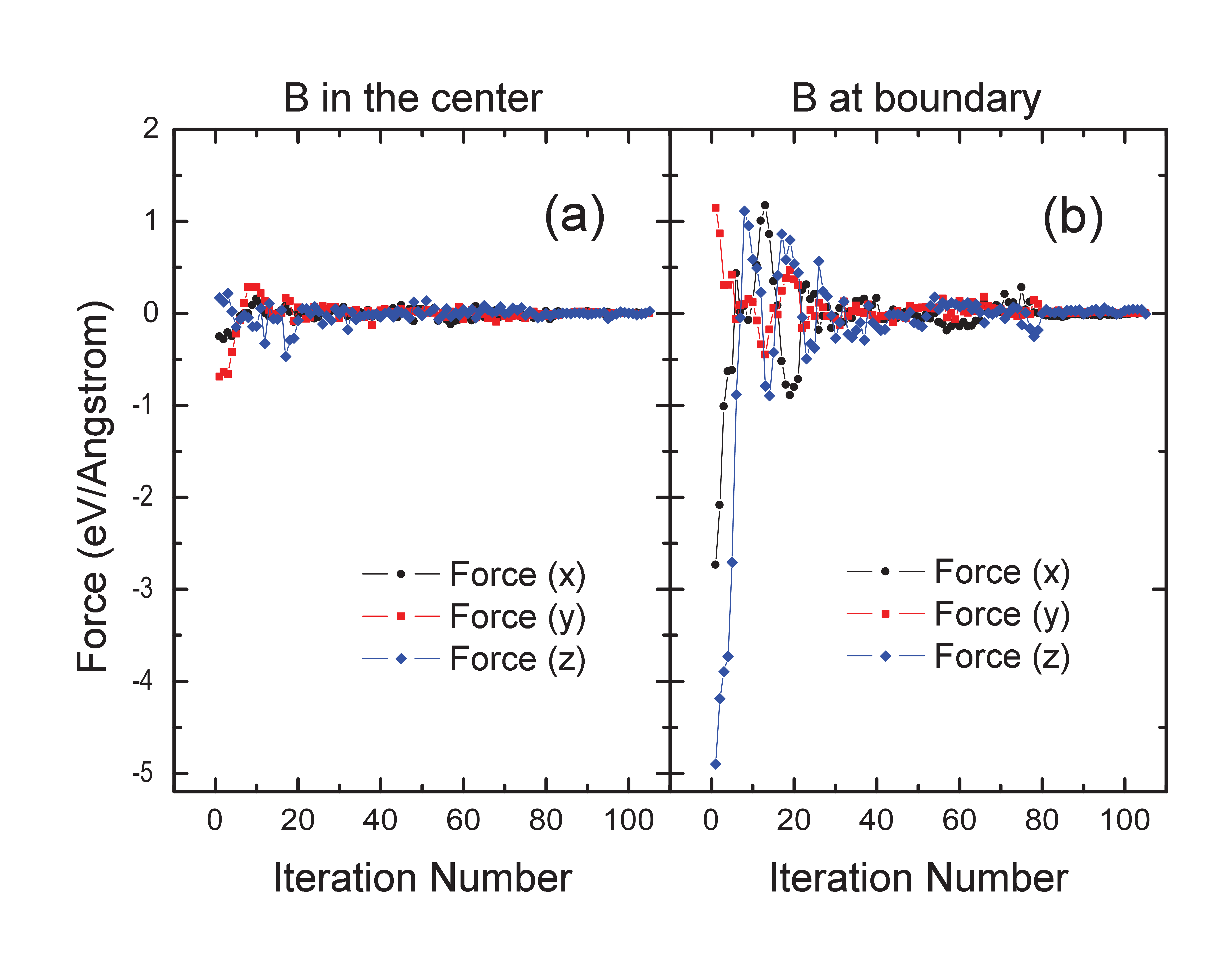}
  \end{center}
  \caption{(color online) The force (x,y,z directions) acting on the
  boron atoms near the center of the nanotube (a) and near the boundary
  of the nanotube (b).}
  \label{fig:BoronForce}
\end{figure}

The convergence history of energy per atom and the convergence history of
the maximum force with respect to the iteration number
in the geometry optimization procedure are plotted in
Fig.~\ref{fig:EnergyForce} (a) and (b), respectively.
In Fig.~\ref{fig:EnergyForce} (a), the energy per atom at the last iteration
step is set to zero. The energy per atom converges rapidly from 0.05 eV to 0.005
eV during the first 16 steps. Correspondingly, in
Fig.~\ref{fig:EnergyForce} (b),
the maximum force converges rapidly during the first few steps. This is
mainly because the initial positions of the hydrogen and boron atoms near
the end of the nanotube are not far from the equilibrium value.  After
the hydrogen and boron atoms at the boundary are relaxed to more
reasonable positions, the maximum force begin to decrease slowly but
with some oscillations.  In order to illustrate more clearly the origin
of the oscillation, we show the forces of boron atoms in
Fig.~\ref{fig:BoronForce}.  Fig.~\ref{fig:BoronForce}(a) and
Fig.~\ref{fig:BoronForce}(b) show the forces of the boron atoms near the
center of the nanotube and near the boundary of the nanotube,
respectively.  We find that the forces acting
on the boron atoms near the center of the nanotube are much smaller than
those near the boundary.  This is mainly due to the fact that
the atomic configuration near the center of the nanotube is close to the
bulk configuration.   The magnitude of the force acting on the atoms
near the boundary is much larger, and is more difficult to convergence
in the numerical optimization.

\section{Conclusion}

In this paper, we generalized the recently developed pole expansion and
selected inversion technique (PEXSI) for solving finite dimensional
Kohn-Sham equations obtained from an atomic orbital expansion.
We gave expressions for evaluating the electron density, the total
energy, the Helmholtz free energy and the atomic forces
(including both the Hellman-Feynman force and the Pulay
force) without using
eigenvalues and eigenvectors of a Kohn-Sham Hamiltonian. These expressions
are derived from an FOE approximation to the Fermi-Dirac function using
an efficient and accurate pole expansion technique.  
The favorable $\log(\beta\Delta E)$ scaling of the pole
expansion allows us to treat both insulating and metallic systems
efficiently at room temperature or even lower temperature.
The pole expansion only uses selected
elements of the density matrix, energy density matrix and
free energy density matrix.  These selected elements can be obtained
from computing the selected elements of the inverse of a shifted Kohn-Sham
Hamiltonian through the selected inversion technique.  The complexity
of the selected inversion is $\mathcal{O}(N_e)$ for quasi-1D systems
such as nanorods, nanotubes and nanowires, $\mathcal{O}(N_e^{3/2})$ for
quasi-2D systems such as graphene and surfaces, and $\mathcal{O}(N_e^2)$
for 3D bulk systems. It compares favorably to the complexity of
diagonalization, which is $\mathcal{O}(N_e^3)$.  We reported the
performance achieved by comparing the efficiency of PEXSI
with that of diagonalization on two types of nanotubes. The linear
scaling behavior of PEXSI with respect to the number of atoms is clear when
the number of atoms in these quasi-1D systems is larger than a few
hundreds. For quasi-2D and quasi-3D systems, we expect the crossover
point over which PEXSI exhibits $\mathcal{O}(N_e^{3/2})$
and $\mathcal{O}(N_e^2)$ scaling to be much larger. However, based
on the experiments presented here, PEXSI may still be more efficient
than diagonalization (before the crossover point is reached) as long as
the Cholesky factors of the shifted Kohn-Sham Hamiltonian are not completely
dense.

The computational experiments we presented above were performed
with a sequential implementation of the selected inversion
algorithm. For quasi-1D systems such as nanotubes,
the use PEXSI allows us to tackle problems that contain as many
as 10,000 atoms. This cannot be done by using a diagonalization
based approach.  We further demonstrate the applicability of the PEXSI
scheme by performing the geometry optimization of a truncated boron
nitride nanotube with 1024 atoms.  For quasi-2D and 3D systems, a
parallel implementation of the PEXSI, which we are currently working on,
is required to solve problems with that many atoms. We will report the
performance for these large-scale calculations in a future publication.


\noindent{\bf Acknowledgment:}
This work was supported by the Laboratory Directed Research and
Development Program of Lawrence Berkeley National Laboratory under
the U.S. Department of Energy contract number DE-AC02-05CH11231 (L. L.
and C. Y.), and by the Chinese National Natural Science Funds for
Distinguished Young Scholars (L. H.).

\appendix

\section*{Appendix}

Derivation of Eq.~\eqref{eqn:gammapole}:

$\Xi$ is a diagonal matrix, and the pole expansion \eqref{eqn:polerho}
can be applied to each component of $\Xi$ as
\begin{equation}
  f_{\beta}(\Xi-\mu) \approx \Im \sum_{l=1}^{P}
  \frac{\omega^{\rho}_l}{\Xi-(z_l+\mu)I},
  \label{}
\end{equation}
where $I$ is an $N\times N$ identity matrix.  Using
Eq.~\eqref{eqn:compactgamma}, the approximation of the single particle
density matrix using $P$ terms of the pole expansion (still denoted by
$\hat{\gamma}$ to simplify the notation) can be
written as
\begin{equation}
  \begin{split}
  \hat{\gamma}(x,x') &=  \Phi(x) C \Im \sum_{l=1}^{P}
  \frac{\omega^{\rho}_l}{\Xi-z_l I} C^T \Phi^T(x') \\
  & = \Phi(x)  \Im \sum_{l=1}^{P}\frac{\omega^{\rho}_l}{C^{-T}\Xi
  C^{-1} - z_l C^{-T} C^{-1}} \Phi^T(x').
  \end{split}
  \label{}
\end{equation}
Since the generalized eigenvalue problem~\eqref{eqn:nonortho} implies the identity
\begin{equation}
  C^T H C = \Xi, \quad C^T S C = I,
  \label{eqn:CHSidentity}
\end{equation}
the single particle density matrix takes the form
\begin{equation}
    \hat{\gamma}(x,x') = \Phi(x) \Im
    \sum_{l=1}^{P}\frac{\omega^{\rho}_l}{H - (z_l+\mu) S} \Phi^T(x')
\end{equation}
which is Eq.~\eqref{eqn:gammapole}.

Derivation of Eq.~\eqref{eqn:Helmholtzpole}:

The first term in the Helmholtz free energy functional is
\begin{equation}
  \begin{split}
    \Tr[f_{\beta}^{\mc{F}}(\Xi-\mu)]
    &= \Tr[Cf_{\beta}^{\mc{F}}(\Xi-\mu)C^T C^{-T}C^{-1}] \\
    &\equiv
    \Tr[\Gamma^{\mc{F}}S].
  \end{split}
  \label{eqn:introgammaF}
\end{equation}
The second equal sign in Eq.~\eqref{eqn:introgammaF} defines the 
free energy density matrix $\Gamma^{\mc{F}}$, which can be evaluated
using  the pole expansion~\eqref{eqn:poleHelmholtz}  as
\begin{equation}
  \begin{split}
   \Gamma^{\mc{F}} &= C\Im \sum_{l=1}^{P}
   \frac{\omega^{\mc{F}}_l}{\Xi-z_l I}C^{T}\\
  &= \Im \sum_{l=1}^{P}
  \frac{\omega^{\mc{F}}_l}{C^{-T}HC^{-1} -z_l C^{-T}C^{-1}} \\
  &= \Im \sum_{l=1}^{P}
  \frac{\omega^{\mc{F}}_l}{H -z_l S},
  \end{split}
  \label{eqn:FreePole}
\end{equation}
which is Eq.~\eqref{eqn:Helmholtzpole}.

Derivation of Eq.~\eqref{eqn:forcepole}:

The atomic force is in general given by the derivative of the Helmholtz
free energy $\mc{F}_{\tot}$ with respect to the atomic positions.  Since the
free energy is minimized with respect to $\{\psi_{i}\}$,$\{f_{i}\}$ at
each atomic configuration $\{R_{I}\}$, all the terms in $\mc{F}_{\tot}$
that do not explicitly depend on $R_{I}$ will not contribute to the
atomic force $F_{I}$. In particular, the double counting terms $-\frac12 \iint
\frac{\hat{\rho}(x) \hat{\rho}(y)}{\abs{x-y}} \ud x \ud y +
E_{\xc}[\hat{\rho}] - \int V_{\xc}[\hat{\rho}](x) \hat{\rho}(x) \ud x$
do not contribute to the atomic force.  Therefore
\begin{equation}
	F_I = -\frac{d}{d R_I} \mc{F}_{\tot}=-\frac{\partial}{\partial R_I}
	\mc{F}_{\tot}.
\end{equation}
Using the representation of the Helmholtz free energy in
Eq.~\eqref{eqn:Helmholtz}, and the fact that
\begin{equation}
  (f_{\beta}^{\mc{F}})'(z) = f_{\beta}(z), \qquad N_e=\Tr\left[
  f_{\beta}(\Xi-\mu)
  \right],
  \label{}
\end{equation}
it can be derived that
\begin{equation}
  \begin{split}
    F_I & = -\frac{\partial}{\partial R_I} \mc{F}_{\tot}
    = -\frac{\partial}{\partial R_I} \left(\Tr[f_{\beta}^{\mc{F}}(\Xi-\mu)]
		+\mu N_e\right)\\
  & = -\Tr\left[(f_{\beta}^{\mc{F}})'(\Xi-\mu) \left(\frac{\partial
  \Xi}{\partial R_I}-\frac{\partial \mu}{\partial R_I}\right)\right]
	- N_e \frac{\partial \mu}{\partial R_I}\\
   &= -\Tr\left[f_{\beta}(\Xi-\mu) \frac{\partial \Xi}{\partial R_I}\right]
	 -  \frac{\partial \mu}{\partial R_I}\left( N_e - \Tr\left[
   f_{\beta}(\Xi-\mu)
   \right] \right)
   \\
  & = -\Tr\left[f_{\beta}(\Xi-\mu) C^T \frac{\partial H}{\partial R_I} C \right]
  -\Tr\left[f_{\beta}(\Xi-\mu) \frac{\partial C^T}{\partial R_I} H C   \right]
  \\
  &\quad -\Tr\left[f_{\beta}(\Xi-\mu) C^T H \frac{\partial C}{\partial R_I}\right] \\
  & =-\Tr\left[\Gamma \frac{\partial H}{\partial R_I} \right]
  -\Tr\left[f_{\beta}(\Xi-\mu) \frac{\partial C^T}{\partial R_I} H C
  \right]\\
  &\quad -\Tr\left[f_{\beta}(\Xi-\mu) C^T H \frac{\partial C}{\partial R_I}\right] \\
  \end{split}
  \label{eqn:forcederive}
\end{equation}
The second and the third terms in Eq.~\eqref{eqn:forcederive} come from the
nonorthogonality of the basis functions and should be further
simplified.  We have
\begin{equation}
  \begin{split}
    & \Tr\left[f_{\beta}(\Xi-\mu) \frac{\partial C^T}{\partial R_I} H C \right]
    + \Tr\left[f_{\beta}(\Xi-\mu) C^T H \frac{\partial C}{\partial R_I}\right] \\
    = & \Tr\left[ (C^{-T} C^{-1}) [C (C^T H C) f_{\beta}(\Xi-\mu) C^T] (C^{-T}
  C^{-1})
  C \frac{\partial C^T}{\partial R_I} \right] \\
  &+ \Tr\left[  C^{-T} C^{-1} [C f_{\beta}(\Xi-\mu) (C^T H C) C^T] C^{-T}
  C^{-1}\frac{\partial C}{\partial R_I} C^T\right] \\
  \equiv &
  \Tr\left[ (C \Xi f_{\beta}(\Xi-\mu) C^T) \left(S C \frac{\partial
  C^T}{\partial R_I} S + S \frac{\partial C}{\partial R_I} C^T S\right)
  \right].
  \end{split}
  \label{eqn:gammaEderive}
\end{equation}
Define the energy density matrix as in Eq.~\eqref{eqn:gammaE},
and Eq.~\eqref{eqn:gammaEderive} can be simplified as
\begin{equation}
  \begin{split}
  &\Tr\left[ \Gamma^E S \left(C \frac{\partial C^T}{\partial R_I}
  + \frac{\partial C}{\partial R_I} C^T\right) S \right]\\
  =&
    \Tr\left[ \Gamma^E S \frac{\partial S^{-1}}{\partial R_I} S
  \right] =  -\Tr\left[ \Gamma^E \frac{\partial S}{\partial R_I} \right]
  \end{split}
  \label{eqn:gammaEderive1}
\end{equation}
Combining Eq.~\eqref{eqn:gammaEderive1} and Eq.~\eqref{eqn:forcederive},
we have
\begin{equation}
  F_{I} = -\frac{\partial \mc{F}}{\partial R_I} =
  -\Tr\left[ \Gamma \frac{\partial H}{\partial R_I} \right]
      +\Tr\left[ \Gamma^E \frac{\partial S}{\partial R_I} \right].
  \label{}
\end{equation}
which proves Eq.~\eqref{eqn:forcepole}.



\end{document}